\documentclass[twocolumn,apj,iop,tighten]{emulateapj2}

\usepackage{float}
\usepackage{graphicx}
\usepackage{xfrac}
\usepackage{adjustbox}
\usepackage{upgreek}

\usepackage{lineno}

 \usepackage{natbib}
\usepackage{xcolor}

\usepackage[ colorlinks=true, 
             urlcolor=blue,
             filecolor=blue,
             linkcolor=red,
             citecolor=blue,
             pdftitle={pdf title},
             pdfauthor={pdf author},
             pagebackref, 
             bookmarks,
             bookmarksopen=true]{hyperref}

\usepackage{xspace}
\usepackage{microtype}
\usepackage{amsmath}
\usepackage{bm}
\usepackage{amssymb}
\usepackage{xfrac}

\usepackage{enumitem}

\usepackage{multirow}
\usepackage{makecell}
\usepackage{rotating}	

\newcommand{\corrfunc}{\texttt{Corrfunc}\xspace}

\newcommand{\chisqred}{\ensuremath{\chi^2_{\nu}}\xspace}

\newcommand{\degsq}{\ensuremath{\mathrm{deg}^2}\xspace}

\newcommand{\Gpch}{\ensuremath{h^{-1}\ {\rm Gpc}}\xspace}

\newcommand{\idl}{\texttt{IDL}\xspace}

\newcommand{\kcorrect}{\texttt{kcorrect}\xspace}

\newcommand{\logm}{\ensuremath{\log(M_*/\msun)}\xspace}

\newcommand{\mchar}{\ensuremath{M_{\rm char}}\xspace}

\newcommand{\msun}{\ensuremath{{\rm M}_\odot}\xspace}
\newcommand{\msunh}{\ensuremath{h^{-1} \msun}\xspace}
\newcommand{\mvir}{\ensuremath{M_{\rm vir}}\xspace}
\newcommand{\Mpch}{\ensuremath{h^{-1}\ {\rm Mpc}}\xspace}

\newcommand{\neff}{\ensuremath{n_{\rm eff}}\xspace}

\newcommand{\pimax}{\ensuremath{\pi_{\rm max}}\xspace}

\newcommand{\rp}{\ensuremath{r_{\rm p}}\xspace}
\newcommand{\sigmalos}{\ensuremath{\sigma_{\rm LOS}}\xspace}
\newcommand{\sigmamag}{\ensuremath{\sigma_{\rm mag}}\xspace}

\newcommand{\vcirc}{\ensuremath{v_{\rm circ}}\xspace}
\newcommand{\vmax}{\ensuremath{v_{\rm max}}\xspace}

\newcommand{\vpeak}{\ensuremath{v_{\rm peak}}\xspace}

\newcommand{\wprp}{\ensuremath{\omega_{\rm p}(r_{\rm p})}\xspace}

\newcommand{\xir}{\ensuremath{\xi(r)}\xspace}

\newcommand{\zacc}{\ensuremath{z_{\rm acc}}\xspace}
\newcommand{\zchar}{\ensuremath{z_{\rm char}}\xspace}
\newcommand{\zform}{\ensuremath{z_{\rm form}}\xspace}
\newcommand{\zmax}{\ensuremath{z_{\rm max}}\xspace}
\newcommand{\zmin}{\ensuremath{z_{\rm min}}\xspace}

\newcommand{\zphot}{\ensuremath{z_{\rm phot}}\xspace}

\newcommand{\zsim}{\ensuremath{z_{\rm sim}}\xspace}
\newcommand{\zstarve}{\ensuremath{z_{\rm starve}}\xspace}

\begin{document}
\title{The galaxy--halo connection of DESI luminous red galaxies with subhalo abundance matching}
\shorttitle{DESI LRG--halo connection with SHAM}
\shortauthors{Berti et al.}

\author{
Angela M.\ Berti\altaffilmark{1},
Kyle S.\ Dawson\altaffilmark{1},
Wilber Dominguez\altaffilmark{2}
}
	
\altaffiltext{1}{Department of Physics \& Astronomy, University of Utah, Salt Lake City, UT 84112, USA}
\altaffiltext{2}{Department of Physics \& Astronomy, Swarthmore College, 500 College Ave, Swarthmore, PA 19081, USA}

\begin{abstract}
We use subhalo abundance and age distribution matching to create magnitude-limited mock galaxy catalogs at $z\sim0.43$, 0.52, and 0.63 with $z$-band and 3.4 micron $W1$-band absolute magnitudes and ${r-z}$ and ${r-W1}$ colors. From these magnitude-limited mocks we select mock luminous red galaxy (LRG) samples according to the $(r-z)$-based (optical) and $(r-W1)$-based (infrared) selection criteria for the LRG sample of the Dark Energy Spectroscopic Instrument (DESI) Survey.
Our models reproduce the number densities, luminosity functions, color distributions, and projected clustering of the DESI Legacy Surveys that are the basis for DESI LRG target selection. We predict the halo occupation statistics of both optical and IR DESI LRGs at fixed cosmology, and assess the differences between the two LRG samples. We find that IR-based SHAM modeling represents the differences between the optical and IR LRG populations better than using the $z$-band, and that age distribution matching overpredicts the clustering of LRGs, implying that galaxy color is uncorrelated with halo age in the LRG regime.
Both the optical and IR DESI LRG target selections exclude some of the most luminous galaxies that would appear to be LRGs based on their position on the red sequence in optical color--magnitude space. Both selections also yield populations with a non-trivial LRG--halo connection that does not reach unity for the most massive halos. We find the IR selection achieves greater completeness ($\gtrsim 90\%$) than the optical selection across all redshift bins studied.
\end{abstract}

\section{Introduction}\label{sec:intro}
The Dark Energy Spectroscopic Instrument \citep[DESI;][]{desi_collab16a, desi_collab16b} survey is a spectroscopic galaxy redshift survey of unprecedented scale that will classify tens of millions of galaxies in four target classes over $\sim14,000$ square degrees at the lowest redshifts of this program. DESI's bright galaxy survey \citep[BGS;][]{hahn_etal22} extends to $z\sim0.4$, while the the luminous red galaxy \citep[LRG;][]{zhou_etal22} sample will reach to $z\sim1$ and cover 20 times the volume of the BGS\footnote{DESI will observe emission line galaxies (ELGs) to $z\sim1.6$, and quasars to $z\sim3.5$.}.

The DESI target samples are optimized for precision measurements of cosmological parameters. However, DESI also offers novel opportunities to study galaxy evolution and the high-mass end of the stellar-to-halo mass relation (SHMR), provided that sample selection effects are well understood.

Unlike the magnitude-limited BGS sample of relatively nearby galaxies, the DESI LRG sample is selected with a comparatively complex set of magnitude and color cuts, creating incompleteness that may depend on any combination of galaxy color, stellar mass, and redshift. However, the LRG sample covers a volume 20 times larger than that of the BGS sample, and will contain a much higher number density of spectroscopic redshifts at ${0.4 < z < 1}$ than any previous spectroscopic galaxy redshift survey, making it a good sample for statistical studies with negligible sample and cosmic variance uncertainties. The rarity and associated low number density of massive galaxies $(\sim2\times10^{-5}\ {\rm Mpc}^{-3}$ for $\logm > 11.5)$ means that such large volumes are essential for obtaining sample sizes large enough for statistically significant measurements of the high-mass end of the galaxy stellar mass function (SMF) and the SHMR of LRGs.

Existing studies of the galaxy--halo relationship for ``DESI-like" LRG samples include halo occupation distribution (HOD) modeling \citep[e.g.,][]{seljak00, berlind_weinberg02, bullock_etal02, berlind_etal03, zheng_etal07, zheng_weinberg07} in \citet{zhou_etal20b}, and semi-analytic modeling in \citet{hernandez-aguayo_etal21} with {\sc Galform} \citep{cole_etal00}, which predicts absolute magnitudes with dust attenuation.

In its simplest form, the HOD model provides a statistical description for how galaxies occupy dark matter halos solely as a function of halo mass.
\citet{zhou_etal20b} fit the projected clustering of ``DESI-like" LRGs measured in five redshift bins from $0.4 < z < 0.9$ with a five-parameter HOD model plus a sixth nuisance parameter to account for photometric redshift uncertainties.
They find similar HOD parameters at $0.4 < z < 0.8$, and statistically significant differences in model parameters for only the highest redshift bin $(0.8 < z < 0.9)$.
While \citet{zhou_etal20b} demonstrate that clustering measurements using photometric redshifts are sufficient to constrain the HOD parameters of ``DESI-like" LRGs, the standard HOD framework they use does not accommodate galaxy populations with halo occupation fractions that do not reach unity for the most massive halos.

\citet{hernandez-aguayo_etal21} apply the {\sc Galform} semi-analytic model (SAM) to the Planck-Millennium cosmological $N$-body simulation \citep{baugh_etal19}.
They then convert predicted absolute magnitudes to apparent magnitudes at various redshift snapshots, enabling the DESI LRG target selection to be applied directly to {\sc Galform} mock galaxy catalogs to select mock LRG samples.
\citet{hernandez-aguayo_etal21} find that the DESI LRG selection criteria exclude a small but important fraction of the most massive galaxies ($\logm > 11.15$).
Consequently, their model predicts that the halo occupation fraction of LRGs does \emph{not} reach unity for the most massive halos, and actually drops with increasing mass, indicative of a non-trivial LRG--halo connection that is not modeled well with a standard HOD.
By comparing the HOD and subhalo mass functions of stellar mass-selected mock galaxies against those of mock LRG samples, \citet{hernandez-aguayo_etal21} show that the DESI LRG selection cuts likely affect the selection of subhalos populated by LRGs, i.e., (sub)halo mass is by itself insufficient to determine whether a subhalo hosts a LRG.

Subhalo abundance matching \citep[SHAM; e.g.,][]{vale_ostriker04, conroy_etal06, behroozi_etal10, reddick_etal13} is an empirical technique for assigning galaxies to dark matter halos in numerical $N$-body simulations by assuming a correlation between a galaxy property---usually luminosity or stellar mass---and halo property such as mass or circular velocity.
In the simplest application of SHAM, mock galaxy catalogs are constructed to reproduce the number density and luminosity or stellar mass function (SMF) of a target dataset with a single free parameter to allow scatter in the galaxy property--halo property correlation.

Extensions to the SHAM framework to incorporate dependencies between additional galaxy and halo properties are broadly referred to as conditional abundance matching \citep[CAM; e.g.,][]{hearin_etal14, zentner_etal14}.
In addition to a primary galaxy property--halo property correlation, CAM models assume a correlation between secondary galaxy and halo properties at a fixed value of the primary property.
Age distribution matching \citep{hearin_watson13} is a form of CAM that equates galaxy color or a similar property with a proxy for the age of dark matter halos.
Unlike the standard HOD framework, in which the statistical relationship between galaxies and dark matter halos is solely a function of stellar and halo mass, CAM can naturally accommodate galaxy assembly bias, the dependence of galaxy properties (besides stellar mass) on the mass accretion history of their host halos \citep[e.g.,][and references therein]{zentner_etal14, wechsler_tinker18}.

The SHAM framework has been used to study the dependence of sample completeness on redshift and stellar mass for LRGs from the Baryon Oscillation Spectroscopic Survey \citep[BOSS;][]{dawson_etal13}, which obtained spectroscopic redshifts of 1.5 million galaxies with $\logm > 11$ to $z\sim0.7$.
BOSS contains two color- and magnitude-selected samples of massive galaxies \citep[$\logm > 11$;][]{reid_etal16}:\ the LOWZ sample of LRGs at $0.15 < z < 0.43$, and the approximately stellar mass-limited constant mass (CMASS) sample, which includes galaxies of all colors at $0.43 < z < 0.8$.

\citet{saito_etal16} use SHAM to construct $z\sim0.5$ mock galaxy catalogs for the BOSS CMASS sample with and without added assembly bias effects.
They use the Stripe 82 Massive Galaxy Catalog \citep[S82-MGC;][]{bundy_etal15} to replicate the total galaxy SMF above $\logm > 10.5$ over $0.43 < z < 0.7$ and assign galaxies to halos in the MultiDark simulation \citep{riebe_etal13}.
\citet{saito_etal16} find that assembly bias does factor into the galaxy--halo connection for high-mass galaxies, i.e., the SHMR for these galaxies has some dependence on galaxy color, and should not be inferred from the clustering signal without any consideration of color.

\citet{yu_etal22} model LRGs from BOSS and eBOSS \citep{dawson_etal16} at ${0.2 < z < 1.0}$ with a SHAM framework that includes two additional free parameters to account for redshift uncertainty and sample incompleteness.

SHAM has also been used to model ``DESI-like" samples at low redshift.
\citet{safonova_etal21} use SHAM and age distribution matching to create $z\sim0.1$ mock galaxy catalogs representative of the DESI BGS sample with $r$-band luminosities and ${g-r}$ colors.
The low redshift range of the BGS sample allows them to utilize spectroscopic redshifts from the Sloan Digital Sky Survey \citep[SDSS;][]{york_etal00} and Galaxy and Mass Assembly \citep[GAMA;][]{loveday_etal12} project.

In this work, we use SHAM and age distribution matching to create magnitude-limited, mock galaxy catalogs at multiple redshifts within the redshift range of the DESI LRG sample.
Two distinct target selection algorithms were considered for the DESI LRG sample:\ an optical selection that uses $r-z$ color, and an infrared selection that uses $r-W1$\footnote{$W1$ is the 3.4 micron band of the Wide-field Infrared Survey Explorer \citep[WISE;][]{wright_etal10}} color.
We select mock LRG samples from our magnitude-limited mocks based on both the optical and IR DESI target selections that match the number density and two-dimensional color--magnitude space distribution of each DESI LRG target sample. We then predict the clustering signal and halo occupation statistics of these samples as a function of redshift.

Our method is novel approach to modeling DESI LRGs that complements existing SAM and HOD models by utilizing the full photometric samples that are the basis of DESI LRG target selection.
We offset the precision of training data lost to photometric redshift errors by driving down cosmic variance uncertainties with complete photometric samples from an unprecedented survey volume.
Finally, this work offers a comparative study of the samples selected by the optical and IR selection algorithms considered for DESI LRGs.

The structure of this paper is as follows. In \S\ref{sec:data_sims} describe the cosmological simulation and photometric galaxy samples used in this work.
\S\ref{sec:model} describes our modeling procedure and the two-point statistics used to constrain our models.
In \S\ref{sec:predict}, we present the predicted properties of LRG samples,
and in \S\ref{sec:summary} we summarize the conclusions of this work.
Where applicable, we assume the cosmological parameter values of \citet{planck_collab16}:\ $h=0.6777$ and $\Omega_{\rm m}=0.307115$, under the assumption of a flat, $\Lambda$CDM cosmological model.

\section{Simulations and data}\label{sec:data_sims}
In this section we describe the cosmological simulation, halo finder, and associated halo properties used for our models. We also present the data from which we select parent galaxy samples for training our models, as well as the DESI LRG target selection functions.

\subsection{Simulations}\label{subsec:sims}

We use halo catalogs and merger histories obtained with the publicly available \texttt{ROCKSTAR} phase-space temporal halo finder \citep{behroozi_etal13b} for the MultiDark Planck 2 (MDPL2) simulation\footnote{www.cosmosim.org} \citep{klypin_etal16}.
MDPL2 assumes Planck cosmology \citep[$h=0.6777$, $\Omega_{\rm m}=0.307115$;][]{planck_collab16} and evolves $3840^3$ dark matter particles in a 1~\Gpch cubic volume, beginning at $z=120$.
The particle resolution is $1.51\times10^9\ h^{-1}\ \msun$. In total 126 snapshots are available between $z\sim15$ and $z=0$. For this work we use three snapshots at $z=0.425$, 0.523, and 0.628.

The \texttt{ROCKSTAR} halo finder is designed to preserve particle--halo membership and identify accurate halo merger trees across multiple time steps of a simulation.
For MDPL2, \texttt{ROCKSTAR} halo catalogs are mass-complete for halos (including subhalos) above ${\mvir\gtrsim11.4\times10^{11}\ h^{-1}\ \msun}$.

\subsubsection{Subhalo abundance matching}\label{subsubsec:sham}

In SHAM modeling, mock galaxies are assigned to dark matter halos by exploiting the correlation between some galaxy property---usually stellar mass or luminosity---and a halo property such as virial mass or circular velocity. Circular velocity is defined as ${\vcirc(r,z) \equiv \sqrt{GM(<r,z)/r}}$, where $M(<r,z)$ is the enclosed mass within radius $r$ at redshift $z$.
SHAM has been tested with several versions of halo circular velocity. In this work we use \vpeak, the peak value of maximum circular velocity (${\vmax(z) = {\rm max}\{v_{\rm circ}(r,z)\}}$) achieved throughout a halo's entire assembly history.

\subsubsection{Age distribution matching}\label{subsubsec:adm}

\begin{figure}
\centering
\includegraphics[width=\linewidth]{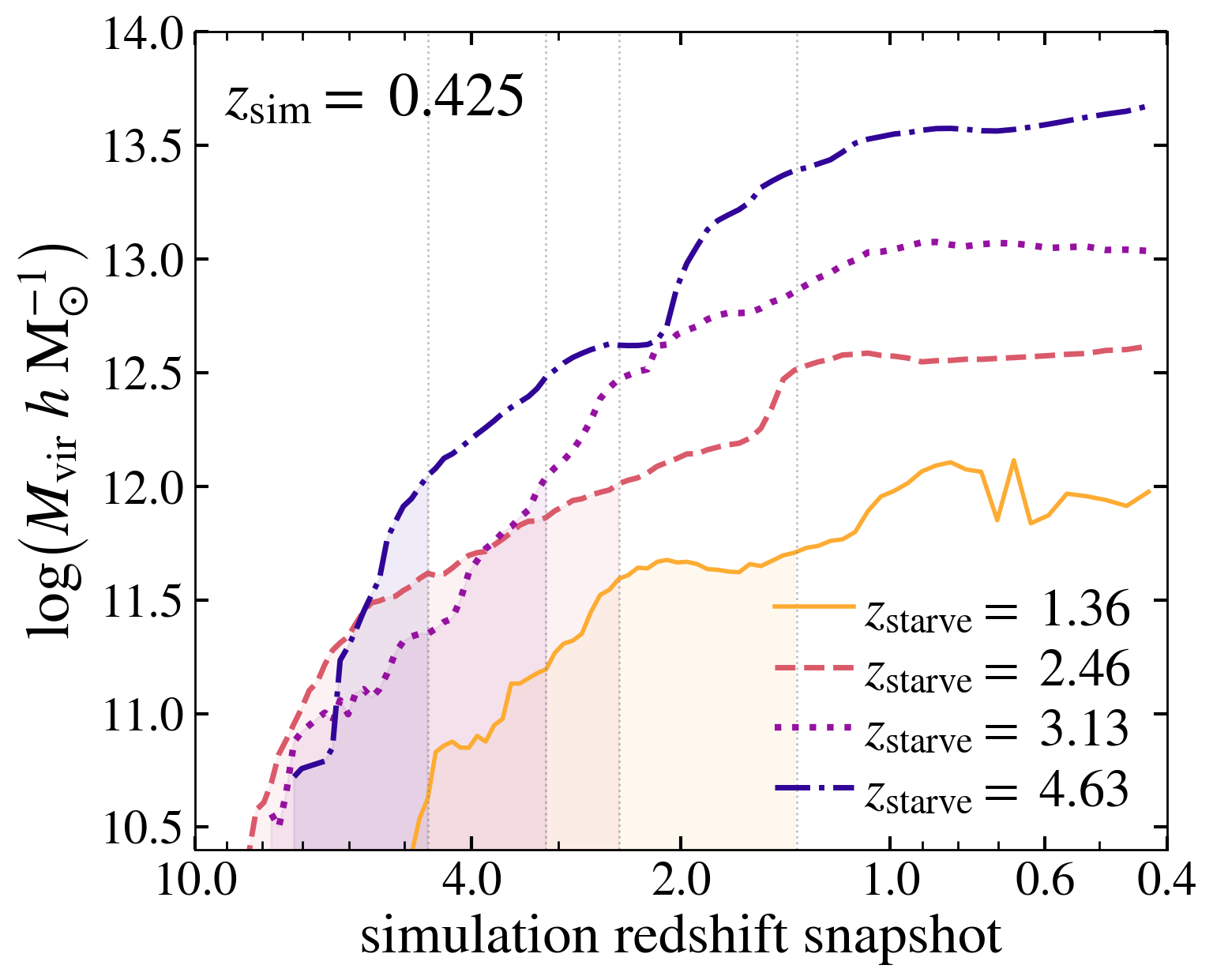}
\caption{
Mass accretion histories and \zstarve values of four randomly selected halos from the $\zsim=0.425$ snapshot of the MDPL2 simulation. The most massive halo (purple dash-dotted line) has the earliest starvation redshift ($\zstarve=4.63$), corresponding to when its mass first reaches the characteristic value ${\mchar=10^{12}\ \msunh}$. The other three halos also reach \mchar (at later redshifts), but in accordance with Eq.~\ref{eq:zstarve} the redshift at which this occurs for each halo is not necessarily the same as its \zstarve value, e.g., the least massive halo (orange solid line) has $\zstarve=1.36$ although its mass doesn't exceed \mchar until $z\lesssim1$. The shaded region below each curve corresponds a halo's mass accretion history before it reaches \zstarve.
}
\label{fig:zstarve_example}
\end{figure}

Age distribution matching assumes a correlation at fixed luminosity between galaxy color and some proxy (at fixed model luminosity) for the age of the halo in which each mock galaxy resides. Redder colors (i.e., older, quenched galaxies) are generally assigned to older halos.
Model colors are assigned at fixed luminosity (in practice in narrow luminosity bins) because the galaxy color distribution is highly dependent on luminosity.

We equate the cumulative distribution $\mathcal{D}_{\rm gal}$ of galaxies of $K$-corrected color $\mathcal{C}$ at fixed absolute magnitude $M_X$ to the cumulative distribution $\mathcal{D}_{\rm halo}$ of halo age proxy $A$ at fixed model absolute magnitude:
\begin{equation}\label{eq:color_assign}
\mathcal{D}_{\rm gal} ( < \mathcal{C}\ |\ M_X )=\mathcal{D}_{\rm halo}( < A\ |\ {\rm model}\ M_X ).
\end{equation}

\noindent In Equation~\ref{eq:color_assign} $(\mathcal{C},M_X) = {(r-z, M_z)}$ OR ${(r-W1, M_{W1})}$ for this work.

Implementations of age distribution matching at $z\sim0$ using spectroscopic galaxy redshifts from SDSS have used halo starvation redshift, \zstarve, for the halo age proxy $A$ in Eq.\ \ref{eq:color_assign} \citep{hearin_watson13, hearin_etal14, safonova_etal21}. In general, \zstarve represents the redshift at which a galaxy loses its supply of cold gas, which leads to the quenching of star formation and the reddening of the galaxy. Multiple physical processes relevant to a halo's assembly history can affect the value of \zstarve for a given halo, which \citet{hearin_watson13} incorporate into the following definition:
\begin{equation}\label{eq:zstarve}
\zstarve \equiv \max\{\zchar,\, \zacc,\, \zform\}.
\end{equation}

\noindent In Equation~\ref{eq:zstarve}:
\begin{itemize}
\item \zchar is either the redshift at which a halo's mass first exceeds some characteristic value, \mchar, or the redshift of the relevant simulation snapshot (\zsim) for halos that never achieve \mchar;
\item \zacc is the redshift at which a subhalo accretes onto a parent halo (for host halos ${\zacc=\zsim}$). \citet{hearin_watson13} follow \citet{behroozi_etal13c}, which defines \zacc as the snapshot after which a subhalo always remains a subhalo. They note that alternative definitions, such as that of \citet{wetzel_etal14}, where \zacc is the snapshot at which a subhalo has been identified as such for two consecutive snapshots, have little impact on their results.
\item \zform is the ``formation" redshift at which a halo transitions from the fast to slow accretion regime.
\end{itemize}
\noindent We use same definition of \zform as \citet{hearin_watson13}, motivated by \citet{wechsler_etal02}:
\begin{equation}\label{eq:zform}
\zform \equiv \frac{c_{\rm vir}}{4.1 a_0} - 1,
\end{equation}

\noindent where ${c_{\rm vir} = R_{\rm vir}/R_{\rm s}}$ is a halo's concentration at the time of observation, indicated by $a_0$. For host halos $a_0$ is the scale factor of the relevant simulation snapshot, while for subhalos $a_0$ is the scale factor at the time of accretion: ${\zacc = 1/{a_0} - 1}$. $R_{\rm vir}$ is the virial radius of a halo, and $R_{\rm s}$ is the NFW scale radius \citep{navarro_etal97}.

We adopt the value of ${\mchar=10^{12}\ \msunh}$ used in \citet{hearin_watson13}, who note that their results are insensitive to the precise value of \mchar used. The empirical and physical motivation for \mchar are described in detail in \S6.3 of \citet{hearin_watson13}. Briefly, there is empirical support for a characteristic halo mass above which star formation is highly inefficient: ${\sim10^{12}\ \msunh}$ is the halo mass at which the SHMR peaks, falling off rapidly at higher halo masses \citep{behroozi_etal13a, yang_etal12, yang_etal13, moster_etal13, watson_conroy13}, and \citet{behroozi_etal13d} have shown that this mass remains essentially constant throughout much of cosmic history.

We compute \zstarve for all halos in our model from the publicly available \texttt{ROCKSTAR} halo merger trees for MDPL2. Figure~\ref{fig:zstarve_example} shows sample halo mass accretion histories and corresponding \zstarve values for four randomly selected halos from the $\zsim=0.425$ snapshot of MDPL2.

\subsection{Photometry and redshift estimates}\label{subsec:photometry}

We use publicly available catalogs from the ninth data release (DR9) of the DESI Legacy Imaging Surveys\footnote{www.legacysurvey.org/dr9} \citep{dey_etal19}. The Legacy Surveys provide optical imaging in the $g$, $r$, and $z$ bands from a combination of three public surveys:\ the DECam Legacy Survey \citep[DECaLS;][]{flaugher_etal15, blum_etal16},
the Beijing-Arizona Sky Survey \citep[BASS;][]{zou_etal17b},
and the Mayall $z$-band Legacy Survey \citep[MzLS;][]{silva_etal16}.
The Legacy Surveys also include four mid-infrared bands from the Wide-field Infrared Survey Explorer \citep[WISE;][]{wright_etal10},
although only the 3.4 micron $W1$-band is relevant for DESI LRG target selection.

In total the Legacy Surveys cover $14,000$ square degrees visible from the northern hemisphere, comprised of two contiguous regions within the northern and southern galactic caps. To avoid effects from systematic differences among data from the three component optical surveys, we limit our study to the approximately 9,000 square degrees covered by DECaLS.

\citet{zhou_etal20b} compute photometric redshifts for the full catalog of DECaLS DR7 objects using the random forest regression machine learning algorithm in \texttt{Scikit-Learn} \citep{pedregosa_etal11} and a ``truth" dataset of spectroscopic and many-band photometric redshifts for objects within DR7.
They quantify the accuracy of their photometric redshifts with the normalized median absolute deviation \citep[NMAD;][]{dahlen_etal13}:\ $\sigma_{\rm NMAD}=1.48 \times {\rm median}( | \Delta z | /(1+z_{\rm spec}))$, where $\Delta z = \zphot - z_{\rm spec}$ and $z_{\rm spec}$ are the redshift truth values used to train the random forest algorithm, and report $\sigma_{\rm NMAD}=0.021$ for LRGs.
Their outlier rate for LRGs is 1\%, where outliers are objects with $| \Delta z | > 0.1\times(1+z_{\rm spec})$.
Additionally, \citet{zhou_etal20b} estimate their redshifts are accurate for objects with apparent $z$-band magnitude $z < 21$, well beyond the $z<20.7$ cut we use to select our target galaxy samples, described in \S\ref{subsec:parent_samples} below.

\subsection{DESI LRG target selection}\label{subsec:lrg_target_select}

The DESI LRG target sample is intended to serve as a cosmological tracer spanning the redshift range $\sim0.4 < z \lesssim 1.0$. The sample lies between the low-redshift Bright Galaxy Survey \citep[BGS;][]{hahn_etal22} tracer sample at $z\lesssim0.4$, and the Emission Line Galaxy \citep[ELG;][]{raichoor_etal22} sample, optimized to trace the density field over the approximate range ${1.0 < z < 1.6}$.

Two different selection algorithms were considered for the DESI LRG sample:\ an optical selection function based on $z$-band magnitude and ${r-z}$ color, and an infrared selection function based on $W1$-band magnitude and ${r-W1}$ color. Both selections are tuned to yield a constant LRG target density of $\sim600$ objects per square degree and a comoving number density around ${5\times10^{-4}\ h^3\ {\rm Mpc}^{-3}}$ at ${0.4 < z < 0.8}$. Both the optical and IR selections were tested in DESI's Survey Validation (SV; DESI et al., in prep.) observations. Based largely on the calibration of $W1$-band imaging, the DESI Main Survey uses exclusively the IR selection. A complete description of DESI LRG target selection is given in \citet{zhou_etal22}. Here we cover the details most relevant for this work.

Due to slight differences in photometry among BASS, MzLS, and DECaLS, the optical and IR DESI LRG target selections use slightly different cuts for the north (BASS and MzLS) and south (DECaLS) galactic caps. As this work uses $g$, $r$, and $z$ magnitudes from DECaLS, we use the corresponding optical LRG target selection cuts \citep{zhou_etal20a}:
\begin{subequations}\label{eq:lrg_opt}
  \begin{align}
   & z - W1 > 0.8\times(r - z) - 0.6\label{eq:lrg_opt_stellar}, \\
   \begin{split}
   & ( (g - W1 > 2.6)\ {\rm AND}\ (g - r > 1.4) )\ {\rm OR} \\
   & \phantom{00} (r - W1 > 1.8)\label{eq:lrg_opt_lowz},
   \end{split} \\
   \begin{split}
   & ( r - z > 0.45\times(z - 16.83) )\ {\rm AND}\ (r - z > 0.7) \\
   & \phantom{00}{\rm AND}\ (r - z > 0.19\times(z - 13.80) )\label{eq:lrg_opt_color-mag},
   \end{split} \\
   & z_{\rm fiber} < 21.5\label{eq:lrg_opt_fiber}.
   \end{align}
\end{subequations}
\noindent The relevant IR LRG target selection cuts for this work are \citep{zhou_etal22}:
\begin{subequations}\label{eq:lrg_ir}
  \begin{align}
   & z - W1 > 0.8 \times (r - z) - 0.6\label{eq:lrg_ir_stellar}, \\
   & ( g - W1 > 2.9)\ {\rm OR}\ (r - W1 > 1.8)\label{eq:lrg_ir_lowz}, \\
   \begin{split}
   & ( ( r - W1 > 1.8 \times (W1 - 17.14) )\ {\rm AND} \\
   & \phantom{00} ( r - W1 > W1 - 16.33 ) )\ {\rm OR}\ ( r - W1 > 3.3 )\label{eq:lrg_ir_color-mag},
  \end{split} \\
   & z_{\rm fiber} < 21.6\label{eq:lrg_ir_fiber}.
  \end{align}
\end{subequations}

Equations~\ref{eq:lrg_opt_stellar} and \ref{eq:lrg_ir_stellar} are designed to reject stars, Eqs.\ \ref{eq:lrg_opt_lowz} and \ref{eq:lrg_ir_lowz} remove blue and low-redshift objects, Eqs.\ \ref{eq:lrg_opt_color-mag} and \ref{eq:lrg_ir_color-mag} are color-dependent magnitude limits that select only the most luminous objects at a given redshift, and $z_{\rm fiber}$ in Eqs.\ \ref{eq:lrg_opt_fiber} and \ref{eq:lrg_ir_fiber} is the expected $z$-band flux within a DESI fiber. All magnitudes in Eqs.\ \ref{eq:lrg_opt} and \ref{eq:lrg_ir} use the AB system, and are corrected for galactic extinction using the relevant \texttt{MW\_TRANSMISSION} values from the Legacy Surveys DR9.

\subsection{Parent galaxy samples}\label{subsec:parent_samples}

\begin{figure}
\centering
\includegraphics[width=\linewidth]{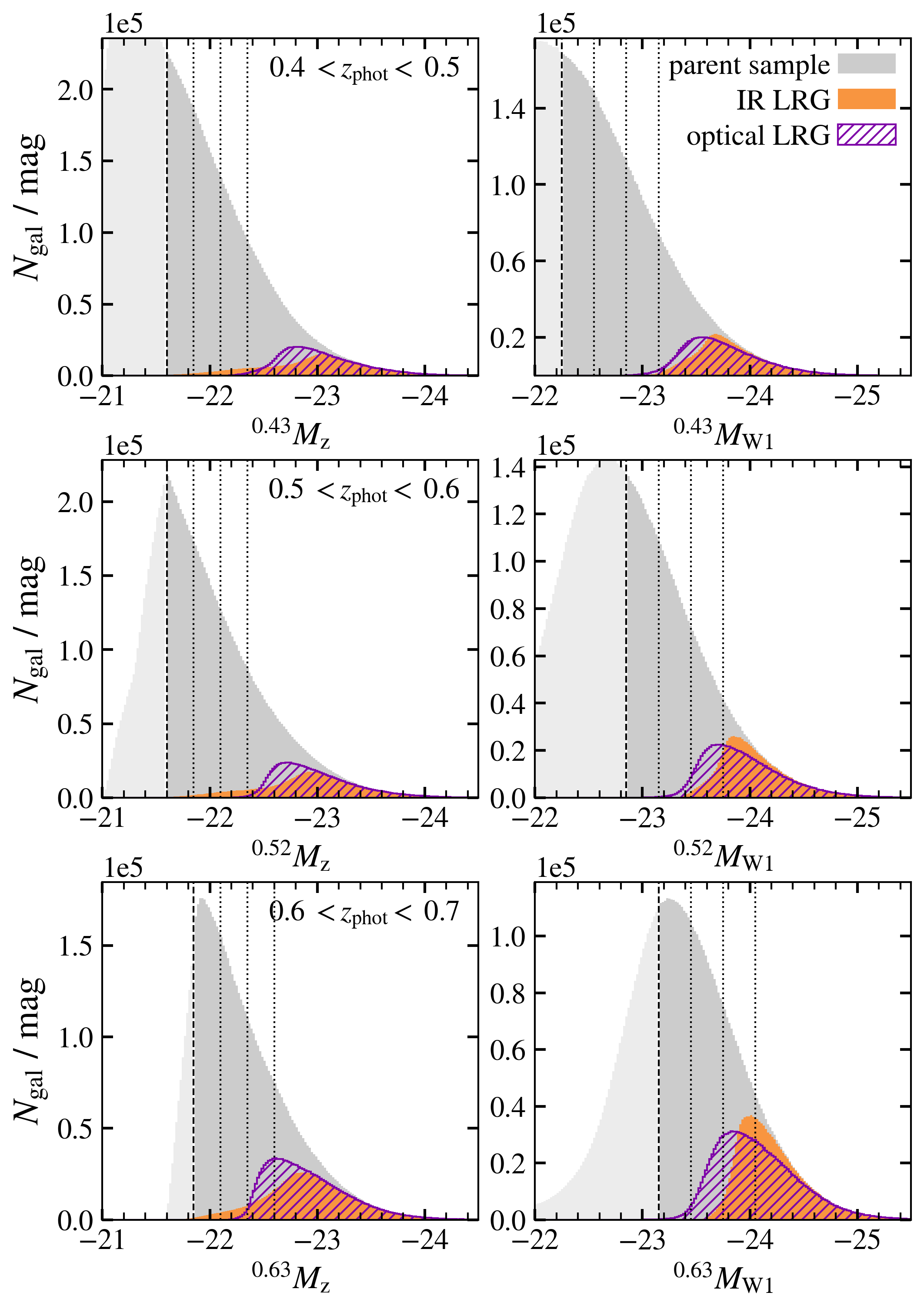}
\caption{Distribution of $K$-corrected $z$-band (left column) and $W1$-band (right column) absolute magnitudes for the three redshift bins in our study:\ ${0.4 < \zphot < 0.5}$ (top), ${0.5 < \zphot < 0.6}$ (middle), and ${0.6 < \zphot < 0.7}$ (bottom). Each panel shows the distribution of all galaxies with $z<20.7$ (solid gray), as well as the distributions of optical (hatched purple; see Eq.~\ref{eq:lrg_opt}) and IR (solid orange; see Eq.~\ref{eq:lrg_ir}) DESI LRG targets. The leftmost dashed black line in each panel is the absolute magnitude cut that defines each absolute magnitude-limited parent galaxy sample for our models. At fainter magnitudes (to the left of the cut in each panel) the DECaLS sample is incomplete. The three dotted black lines in each panel denote the luminosity bins used to constrain model parameters (see \S\ref{subsec:error} and Tables \ref{tab:parent_samples} and \ref{tab:mag_bins}).
}
\label{fig:abs_mag_cuts}
\end{figure}

\begin{figure}
\centering
  \setlength{\tabcolsep}{0pt}
    \begin{tabular}{ c c }
      \includegraphics[width=0.5\linewidth, trim=0 3cm 0 2mm, clip]{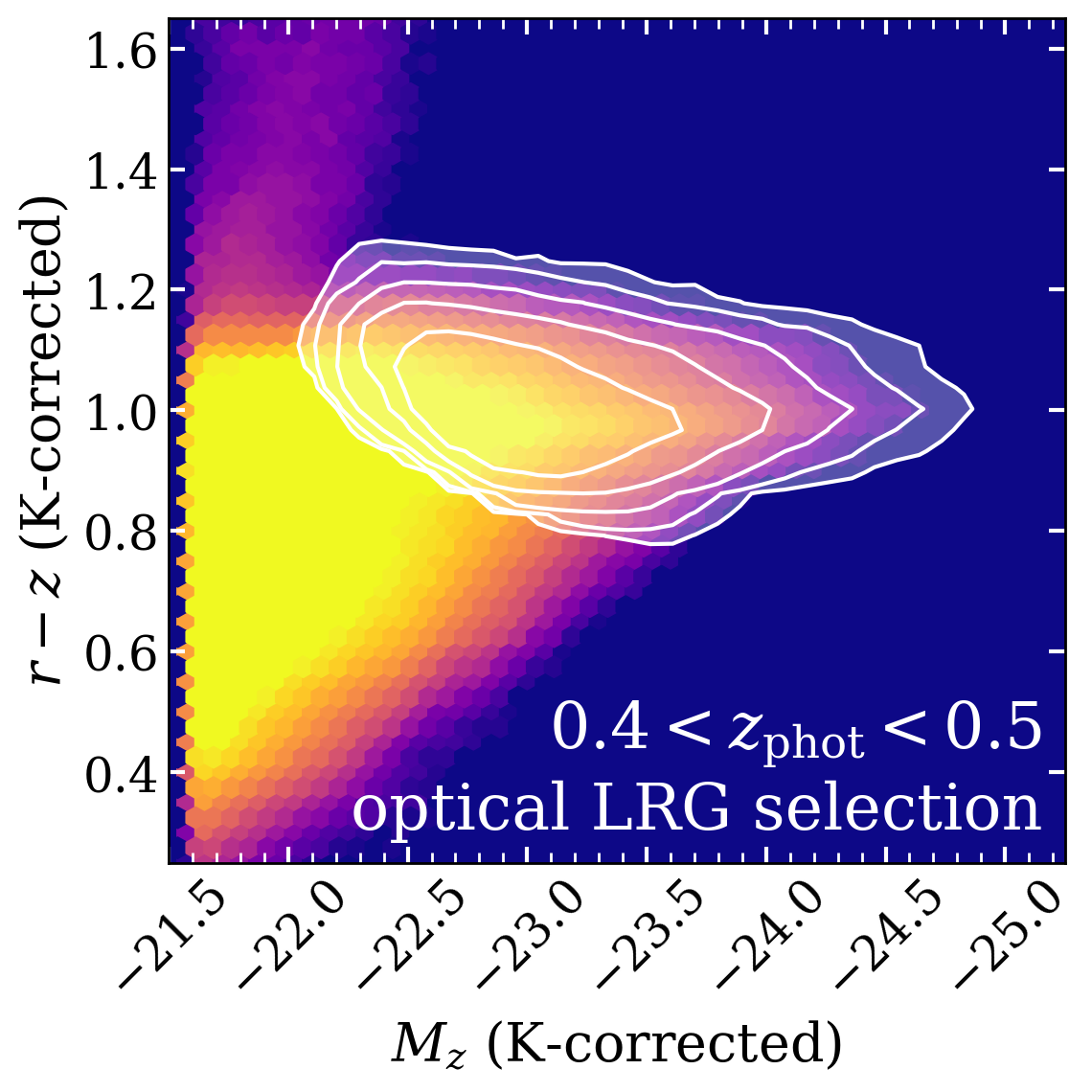} &
      \includegraphics[width=0.5\linewidth, trim=0 2.9cm 0 0, clip]{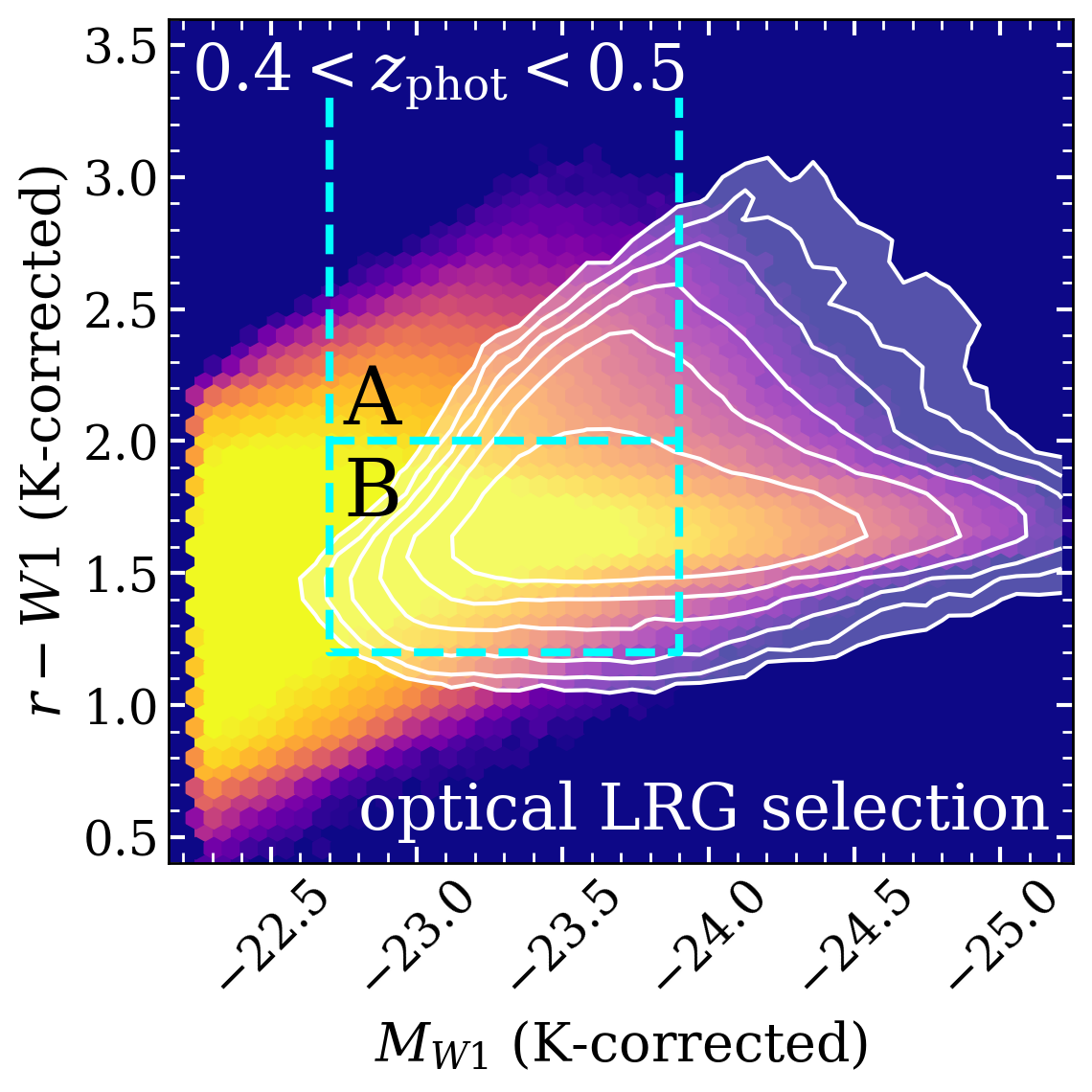} \\
      \includegraphics[width=0.5\linewidth, trim=0 0 0 2mm, clip]{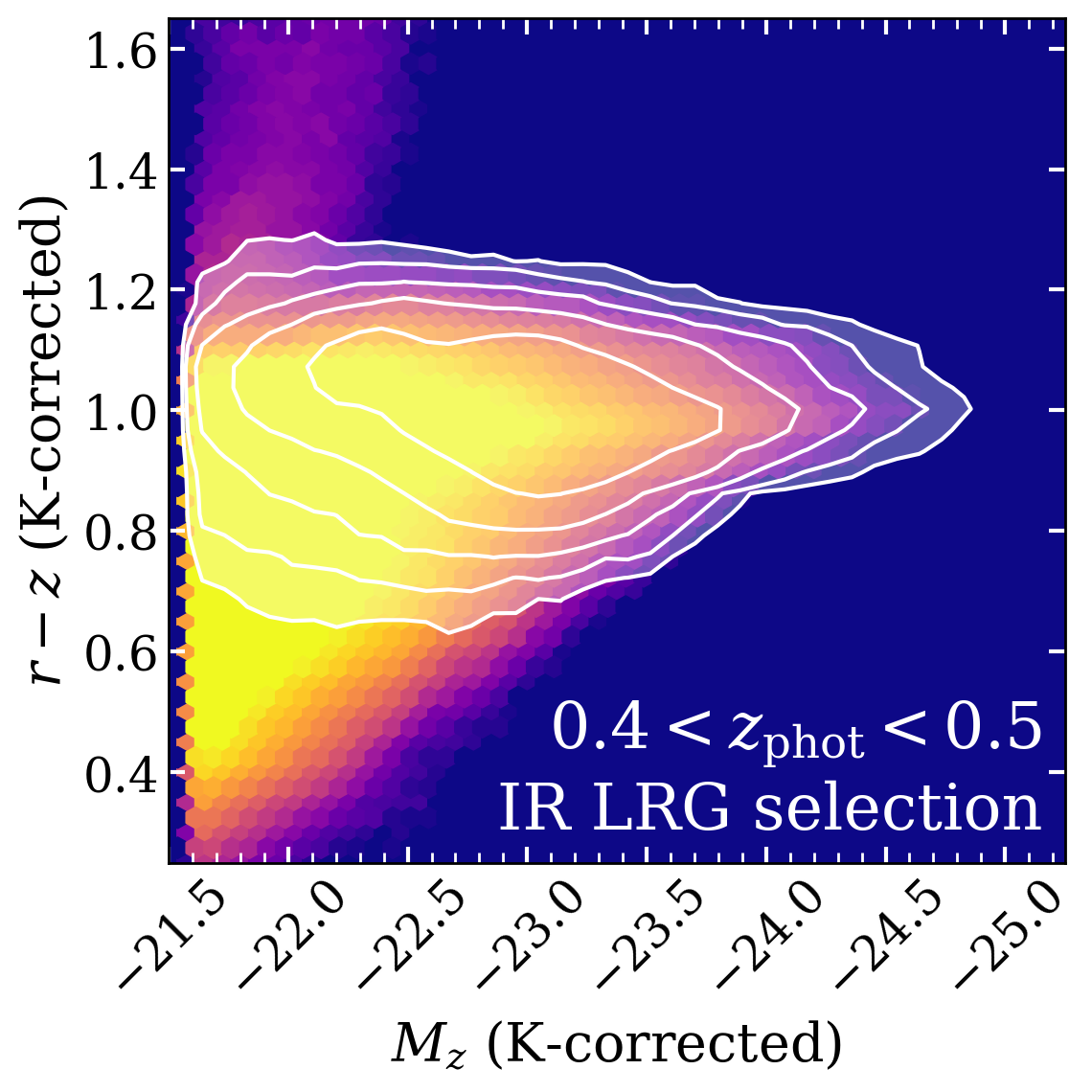} &
      \includegraphics[width=0.5\linewidth, trim=0 0 0 0, clip]{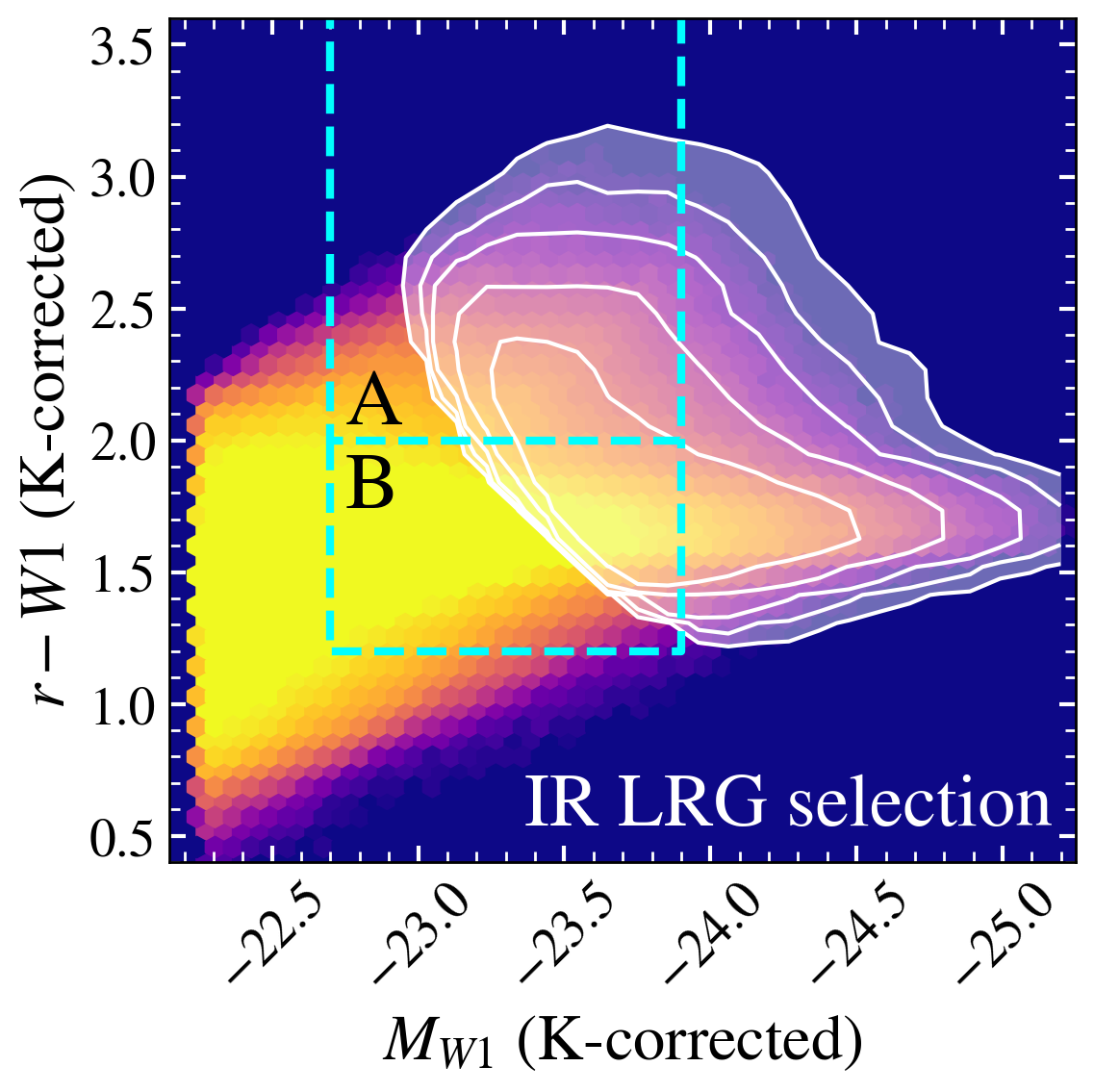} \\
    \end{tabular}
  \caption{
  Color--magnitude diagrams of the parent galaxy samples and DESI LRG target samples (overlaid white contours) in the {$0.4 < \zphot < 0.5$} redshift bin. The left column shows optical ($K$-corrected $r-z$ color versus $M_z$ magnitude), while the right column shown IR ($K$-corrected $r-W1$ color versus $M_{W1}$ magnitude).
  The top (bottom) row shows the distribution of optical (IR) LRGs in each color--magnitude space.
  The boxed regions labeled ``A" and ``B" are referenced in \S\ref{subsec:clust_lrg} below.
 }
  \label{fig:cmd}
\end{figure}

\begin{deluxetable}{ r l }
\tablecaption{
Bitmasks applied to photometry for parent galaxy sample selection.
Additional details at legacysurvey.org/dr9/bitmasks.
\label{tab:masks}
}
\tablehead{
\colhead{\texttt{MASKBIT}} & \colhead{Description}
}
\startdata
{5, 6, 7} & {bad pixel in all of a set of overlapping $g$, $r$, or $z$-band } \\
{} & {images} \\
{8, 9} & bad pixel in a WISE $W1$ or $W2$ bright star mask \\
{11} & {pixel within locus of a radius-magnitude relation for} \\
{} & {Gaia\tablenotemark{a} DR2 stars to $G < 16$} \\
{12} & pixel in a Siena Galaxy Atlas\tablenotemark{b} large galaxy \\
{13} & pixel in a globular cluster \\
\hline \\
\vspace{-1ex} \\
\colhead{\texttt{FITBIT}} & \colhead{Description} \\
\vspace{-1ex} \\
\hline \\
\vspace{-1ex} \\
{6} & source is a medium-bright star \\
{7} & Gaia source\tablenotemark{c} \\
{8} & Tycho-2 star\tablenotemark{d} \\
\enddata
\tablenotetext{a}{\citet{gaia_collab18}}
\tablenotetext{b}{\citet{moustakas_etal21}}
\tablenotetext{c}{\citet{gaia_collab16}}
\tablenotetext{d}{\citet{hog_etal00}}
\end{deluxetable}

\begin{deluxetable*}{ r r r r c c c }
\tablecaption{Properties of parent galaxy samples and corresponding simulation snapshot redshifts (\zsim).
\label{tab:parent_samples}
}
\tablehead{
\colhead{\multirow{2}{*}{Redshift bin}} & \colhead{\multirow{2}{*}{\zsim}} & \colhead{\multirow{2}{*}{Luminosity cut\tablenotemark{a}}} & \colhead{\multirow{2}{*}{$N_{\rm gal}$}} & \colhead{\neff} & \multicolumn{2}{c}{Included fraction of DESI LRG targets} \\
\colhead{} & \colhead{} & \colhead{} & \colhead{} & \colhead{$[\times 10^{-3}\ h^3\ {\rm Mpc}^{-3}]$} & \colhead{Optical selection} & \colhead{IR selection}
}
\startdata
\multirow{2}{*}{$0.4 < \zphot < 0.5$} & \multirow{2}{*}{0.425} & $^{0.43}M_z < -21.60$ & 8,314,309 & 5.60 & 0.992 & 0.982 \\
& & $^{0.43}M_{W1} < -22.25$ & 7,565,153 & 4.33 & 0.992 & 0.992 \\
\vspace{-1ex} \\
\hline
\vspace{-1ex} \\
\multirow{2}{*}{$0.5 < \zphot < 0.6$} & \multirow{2}{*}{0.523} & $^{0.52}M_z < -21.60$ & 7,804,346 & 2.86 & 0.992 & 0.988 \\
& & $^{0.52}M_{W1} < -22.85$ & 4,909,857 & 2.01 & 0.992 & 0.992 \\
\vspace{-1ex} \\
\hline
\vspace{-1ex} \\
\multirow{2}{*}{$0.6 < \zphot < 0.7$} & \multirow{2}{*}{0.628} & $^{0.63}M_z < -21.85$ & 6,548,126 & 1.46 & 0.994 & 0.990 \\
& & $^{0.63}M_{W1} < -23.15$ & 4,758,470 & 1.12 & 0.993 & 0.994 \\
\enddata
\tablenotetext{a}{$K$-correction redshifts are rounded to two decimal places for clarity, e.g., $^{0.43}M_z$ indicates absolute $z$-band magnitudes are $K$-corrected to $\zsim=0.425$.}
\end{deluxetable*}

\begin{deluxetable*}{ r r r r r r }
\tablecaption{Luminosity bins used to constrain the magnitude dependence of model parameters.
\label{tab:mag_bins}
}
\tablehead{
\multicolumn{2}{c}{$0.4 < \zphot < 0.5$} & \multicolumn{2}{c}{$0.5 < \zphot < 0.6$} & \multicolumn{2}{c}{$0.6 < \zphot < 0.7$} \\
\vspace{-1ex} \\
\hline
\vspace{-1ex} \\
\colhead{Luminosity bin} & \colhead{$N_{\rm gal}$} & \colhead{Luminosity bin} & \colhead{$N_{\rm gal}$} & \colhead{Luminosity bin} & \colhead{$N_{\rm gal}$} \\
\vspace{-1ex} \\
\hline
\vspace{-1ex} \\
\multicolumn{6}{c}{$z$-band}
}
\startdata
$-21.60 > {^{0.43}M_z} > -21.85$ & 2,458,920 & $-21.60 > {^{0.52}M_z} > -21.85$ & 2,304,853 & $-21.85 > {^{0.63}M_z} > -22.10$ & 2,040,336 \\
$-21.85 > {^{0.43}M_z} > -22.10$ & 1,880,295 & $-21.85 > {^{0.52}M_z} > -22.10$ & 1,732,820 & $-22.10 > {^{0.63}M_z} > -22.35$ & 1,512,940 \\
$-22.10 > {^{0.43}M_z} > -22.35$ & 1,317,269 & $-22.10 > {^{0.52}M_z} > -22.35$ & 1,205,545 & $-22.35 > {^{0.63}M_z} > -22.60$ & 1,039,573 \\
$ {^{0.43}M_z} <-22.35$ & 1,919,594 & $ {^{0.52}M_z} < -22.35$ & 1,855,014 & $ {^{0.63}M_z} < -22.60$ & 1,401,662 \\
\cutinhead{$W1$-band}
$-22.25 > {^{0.43}M_{W1}} > -22.55$ & 2,324,816 & $-22.85 > {^{0.52}M_{W1}} > -23.15$ & 1,771,554 & $-23.15 > {^{0.63}M_{W1}} > -23.45$ & 1,634,860 \\
$-22.55 > {^{0.43}M_{W1}} > -22.85$ & 1,837,859 & $-23.15 > {^{0.52}M_{W1}} > -23.45$ & 1,245,107 & $-23.45 > {^{0.63}M_{W1}} > -23.75$ & 1,273,814 \\
$-22.85 > {^{0.43}M_{W1}} > -23.15$ & 1,277,192 & $-23.45 > {^{0.52}M_{W1}} > -23.75$ & 754,154 & $-23.75 > {^{0.63}M_{W1}} > -24.05$ & 798,012 \\
$ {^{0.43}M_{W1}} <-23.15$ & 1,565,914 & ${^{0.52}M_{W1}} <-23.75$ & 693,491 & $ {^{0.63}M_{W1}} < -24.05$ & 685,799 \\
\enddata 
\end{deluxetable*}

A primary goal of this work is to create mock galaxy catalogs that are both statistically complete and represent a superset of the color--magnitude space occupied by DESI LRG targets. Selection of DESI LRG targets is based entirely on $g$, $r$, $z$, and $W1$ apparent magnitudes (see \S\ref{subsec:lrg_target_select}), so we would ideally create mock catalogs where every mock galaxy has an apparent magnitude in each of these bands. SHAM, however, exploits the correlation between some physical halo property (e.g., circular velocity) and a physical galaxy property independent of redshift (e.g., luminosity). We therefore train our mock catalogs on galaxy samples that are complete to an \emph{absolute} magnitude threshold that includes all DESI LRG targets (in the relevant redshift bin; see below).

To select suitable parent galaxy samples, we first apply an apparent $z$-band magnitude cut of $z < 20.7$, and take an additional step to remove stars by excluding catalog sources with ${\rm TYPE} = \texttt{PSF}$. We also apply the masks described in Table~\ref{tab:masks}, which are provided with DECaLS DR9, to remove sources affected by bad pixels or contamination from bright stars. Finally, a geometric mask is applied to ensure complete angular coverage by the catalogs of random points provided with DR9 (see \S\ref{subsec:wprp}). The resulting sample contains $\mathcal{O}(10^8)$ galaxies, sufficient to divide it into redshift bins and maintain low statistical error.

We initially tested six redshift bins of width $\Delta\zphot=0.1$ between $\zphot=0.4$ and $\zphot=1.0$, but found that DECaLS photometry is only deep enough to apply our model up to $\zphot \sim 0.7$. At $\zphot \gtrsim 0.7$ the data are incomplete above the absolute magnitude threshold that encompasses DESI LRG targets. We therefore limit our study to three redshift bins of $\Delta\zphot=0.1$ within $0.4 < \zphot < 0.7$.

For each redshift bin we compute $z$- and $W1$-band absolute magnitudes using photometric redshifts to obtain distance moduli.
We $K$-correct absolute magnitudes to the redshift of the relevant simulation snapshot, \zsim (see Table~\ref{tab:parent_samples}) with the \idl package \kcorrect \citep{blanton_roweis07}
using DECam $g$, $r$, and $z$ and WISE $W1$ and $W2$ filter responses.
For each redshift bin we use the simulation snapshot closest to the median \zphot of the data, e.g., galaxies in the ${0.4 < \zphot < 0.5}$ bin are $K$-corrected to $\zsim=0.425$.

The final step in selecting parent galaxy samples is to identify $z$- and $W1$-band absolute magnitude cuts in each redshift bin that yield complete samples which also include the full absolute magnitude range of DESI LRG targets in that bin. Figure~\ref{fig:abs_mag_cuts} shows $K$-corrected absolute magnitude distributions for all $z<20.7$ DECaLS galaxies in each redshift bin. For each bandpass and redshift bin combination, we identify an absolute magnitude cut (dashed black lines in Figure~\ref{fig:abs_mag_cuts}) that eliminates fainter galaxies where DECaLS becomes incomplete, while preserving $\gtrsim99\%$ of DESI LRG targets (solid orange and hatched purple histograms).

Table~\ref{tab:parent_samples} lists the details of each parent galaxy sample, including the relevant absolute magnitude cut, sample size, effective number density, and the included fractions of IR and optical DESI LRG targets. Besides enforcing statistically complete parent samples, these magnitude cuts also eliminate galaxies with larger photometric redshift errors, increasing the accuracy of clustering measurements (see \S\ref{subsec:wprp}).

Figure~\ref{fig:cmd} shows example optical ($r-z$ versus $M_z$) and IR ($r-W1$ versus $M_{W1}$) color--magnitude diagrams of the magnitude-limited parent galaxy samples for the {$0.4 < \zphot < 0.5$} redshift bin. Also shown in Figure~\ref{fig:cmd} are the color--magnitude distributions of optical and IR DESI LRG targets.

\section{Modeling}\label{sec:model}
In this section we describe our modeling procedure and the two-point statistics we use to constrain the model parameters.

\subsection{Projected correlation functions}\label{subsec:wprp}

One goal of this work is to exploit the completeness and enormous volume of DECaLS data, which comes at the expense of the precision clustering measurements achievable with spectroscopic redshifts. \citet{zhou_etal20b} demonstrate the constraining power of the projected correlation function, \wprp, of DECaLS galaxies computed with line-of-sight distances derived from photometric redshifts (they use this statistic to compute HOD parameters of ``DESI-like" LRGs selected from DECaLS DR7). The projected correlation function conveys 3D correlation function, \xir, integrated along the line-of-sight, effectively eliminating the effects of radial distance uncertainty due to photometric redshift errors:
\begin{equation}
\wprp \equiv \int_{-\pimax}^{\pimax}\!\xi(\rp,\pi)\,d\pi = 2\!\int_0^{\pimax}\!\xi(\rp,\pi)\,d\pi,
\end{equation}
 
\noindent where \rp is the projection of $r$ into the plane perpendicular to the line-of-sight separation, $\pi$.

We use the \corrfunc package \citep{sinha_garrison17, sinha_garrison19} to calculate \wprp for both our target galaxy samples and mock catalogs in 19 logarithmic bins between ${\rp>0.1\ \Mpch}$ and ${\rp \lesssim 44\ \Mpch}$. We also compute \wprp in additional bins at ${\rp<0.1\ \Mpch}$ but do not use these measurements for model fitting.

As our data samples are confined to narrow redshift bins of width ${\Delta z_{\rm phot} = 0.1}$, photometric redshift errors will cause some galaxies that belong to a given redshift bin to scatter into an adjacent bin and be excluded from the calculation of \wprp for their true bin. To account for this we adopt the method used by \citet{zhou_etal20b} (see their Figure 8):\ we use the Landy-Szalay estimator \citep{landy_szalay93} for the cross-correlation of two samples, $D_1$ and $D_2$:
\begin{equation}\label{eq:landy-szalay}
\wprp =\!\!\sum_{-\pimax}^{\pimax}\!\!\left(\frac{D_1D_2 - D_1R_2 - D_2R_1}{R_1R_2} + 1 \right),
\end{equation}

\noindent Each term of Eq.\ \ref{eq:landy-szalay} denotes pair counts between two samples, where $D$ and $R$ respectively indicate samples of data (i.e., galaxies) and random points with the same angular and redshift distributions as the corresponding data sample. Here $D_1$ is all galaxies within a given redshift bin: ${\zmin < \zphot < \zmax}$, where \zmin and \zmax are the limits of the bin, while $D_2$ is all galaxies within a wider redshift range defined by ${(\zmin-\pimax) < \zphot < (\zmax+\pimax)}$, where ${\pimax=150~\Mpch}$.
We verify our implementation of this method with \corrfunc by reproducing the projected correlation functions of DECaLS LRGs from \citet{zhou_etal20b} (see their Figure 9) using different clustering code.

DECaLS data include catalogs of random points with the same angular sky coverage and mask information as the survey footprint, which we use to construct our random samples. We use 20 times as many random as data points for each galaxy sample, and draw redshifts for random points from the redshift distribution of the corresponding data sample.

To measure \wprp of our mock catalogs we take advantage of the \corrfunc \texttt{theory} module, which can quickly calculate the autocorrelation function of a sample within a periodic volume using analytic random pair counts. We confirmed that this method produces the expected result by calculating \wprp of several mock catalogs directly from pair counts between mock galaxies and catalogs of random points constructed for the simulation volume.

\subsection{Jackknife error estimation and goodness-of-fit}\label{subsec:error}

To estimate the uncertainty of the \wprp measurements of our target galaxy samples we use \texttt{healpy}\footnote{healpix.sourceforge.net} \citep{gorski_etal05, zonca_etal19} with $N_{\rm side}=6$ to divide the angular sky coverage of each galaxy sample into $N_{\rm jk}$ regions of roughly equal area, suitable for jackknife resampling. We then measure \wprp in each jackknife sample, where each jackknife sample consists of the entire galaxy sample with one jackknife region removed, and compute the covariance matrix as follows:
\begin{equation}\label{eq:cov}
{\rm Cov}_{ij} = \frac{N_{\rm jk}-1}{N_{\rm jk}}
\sum^{N_{\rm jk}}_{\ell=1}\left(\omega^\ell_i-\overline{\omega}_i\right) \left(\omega^\ell_j-\overline{\omega}_j\right),
\end{equation}

\noindent where $\omega^\ell_i$ and $\omega^\ell_j$ are \wprp of the $\ell$th jackknife region for the $i$th and $j$th \rp bins, respectively, and $\overline{\omega}_i$ and $\overline{\omega}_j$ are the mean values of \wprp across all jackknife regions for the $i$th and $j$th \rp bins, respectively.

With the covariance matrix in hand we quantify how successful any instance of our model is at fitting the projected correlation function of the data by computing $\chi^2$ per degree of freedom $(\chisqred)$:
\begin{equation}\label{eq:chisq}
\chisqred=\frac{1}{\nu}\sum_{i=1}^{N_{r_{\rm p}}}
\sum_{j=1}^{N_{r_{\rm p}}}\left(\omega_i-\omega^{\rm mod}_i\right)
\!\left({\rm Cov}^{-1}\right)_{ij}\!\left(\omega_j-\omega^{\rm mod}_j\right)\!,
\end{equation}

\noindent where $N_{\rp}$ is the number of \rp bins used for fitting, $\nu$ is equal to $N_{\rp}$ minus the number of free model parameters, $\omega_i$ and $\omega_j$ are the data \wprp values in the $i$th and $j$th \rp bins, respectively, and $\omega^{\rm mod}_i$ and $\omega^{\rm mod}_j$ are the \wprp values of the relevant mock catalog in the $i$th and $j$th \rp bins, respectively.

\subsection{Luminosity assignment}\label{subsec:luminosity_assign}

\begin{figure*}
\centering
  \includegraphics[width=0.3549\linewidth]{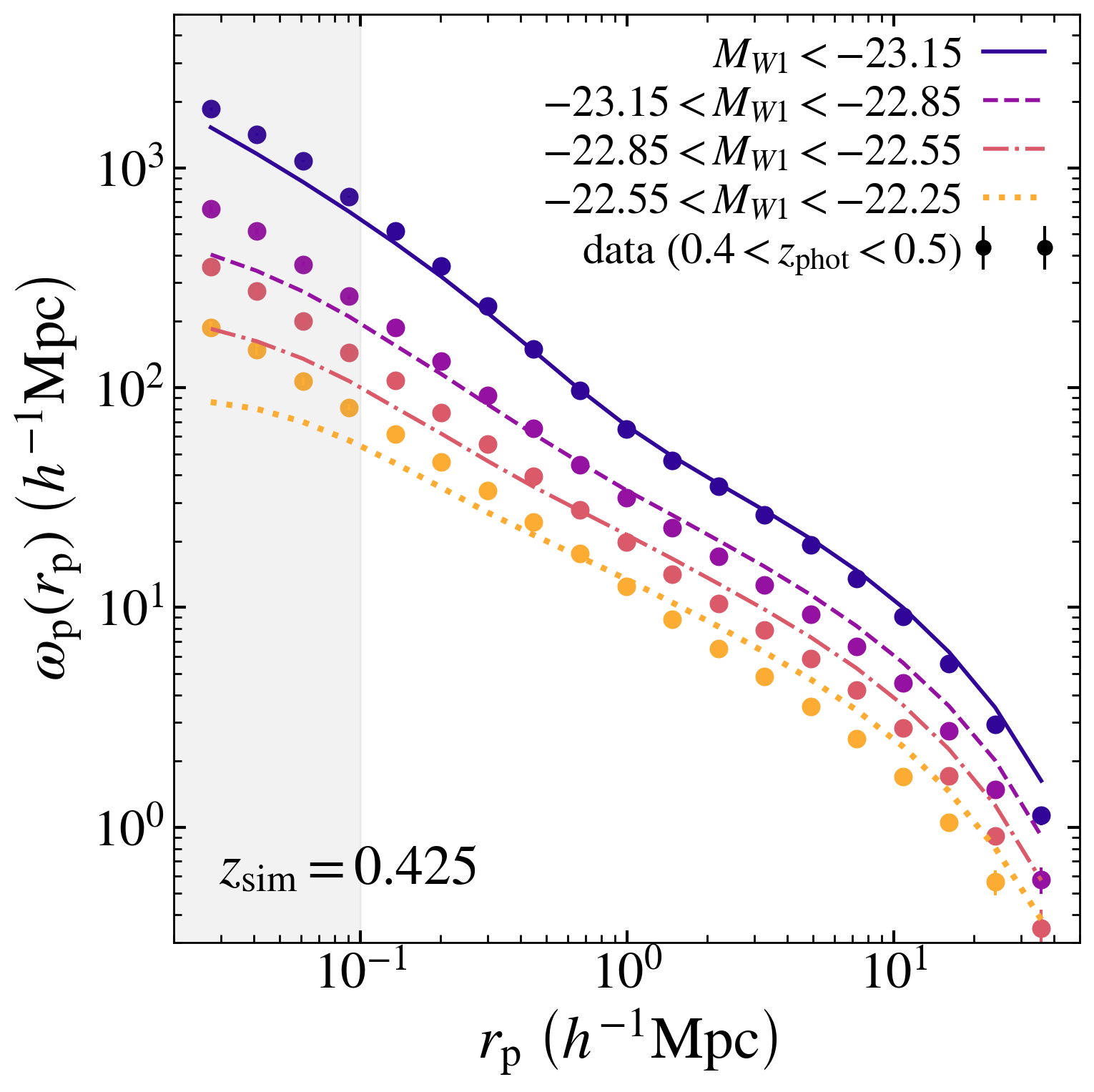}
  \includegraphics[width=0.30\linewidth, trim=3cm 0 0 0, clip]{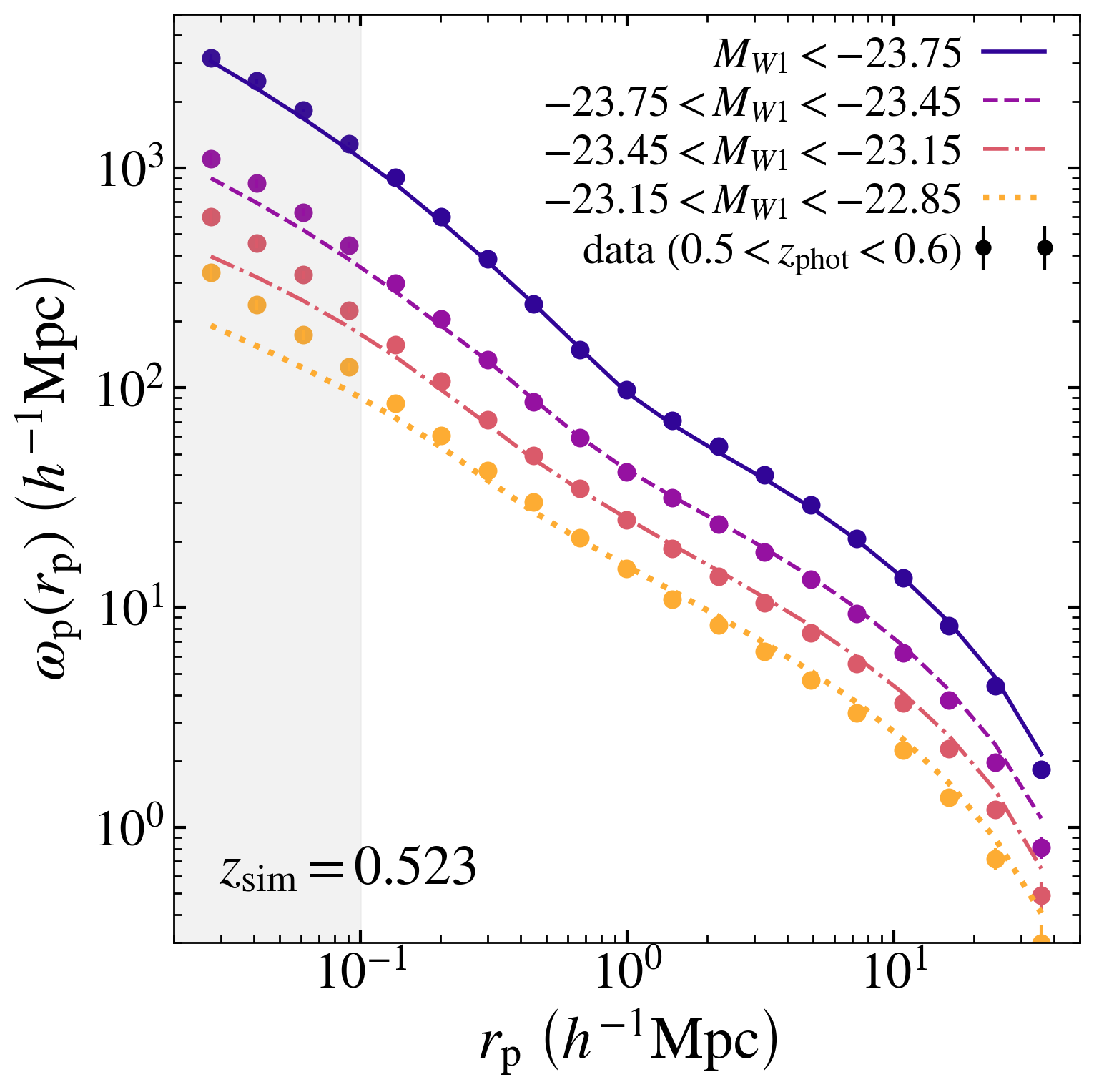}
  \includegraphics[width=0.30\linewidth, trim=3cm 0 0 0, clip]{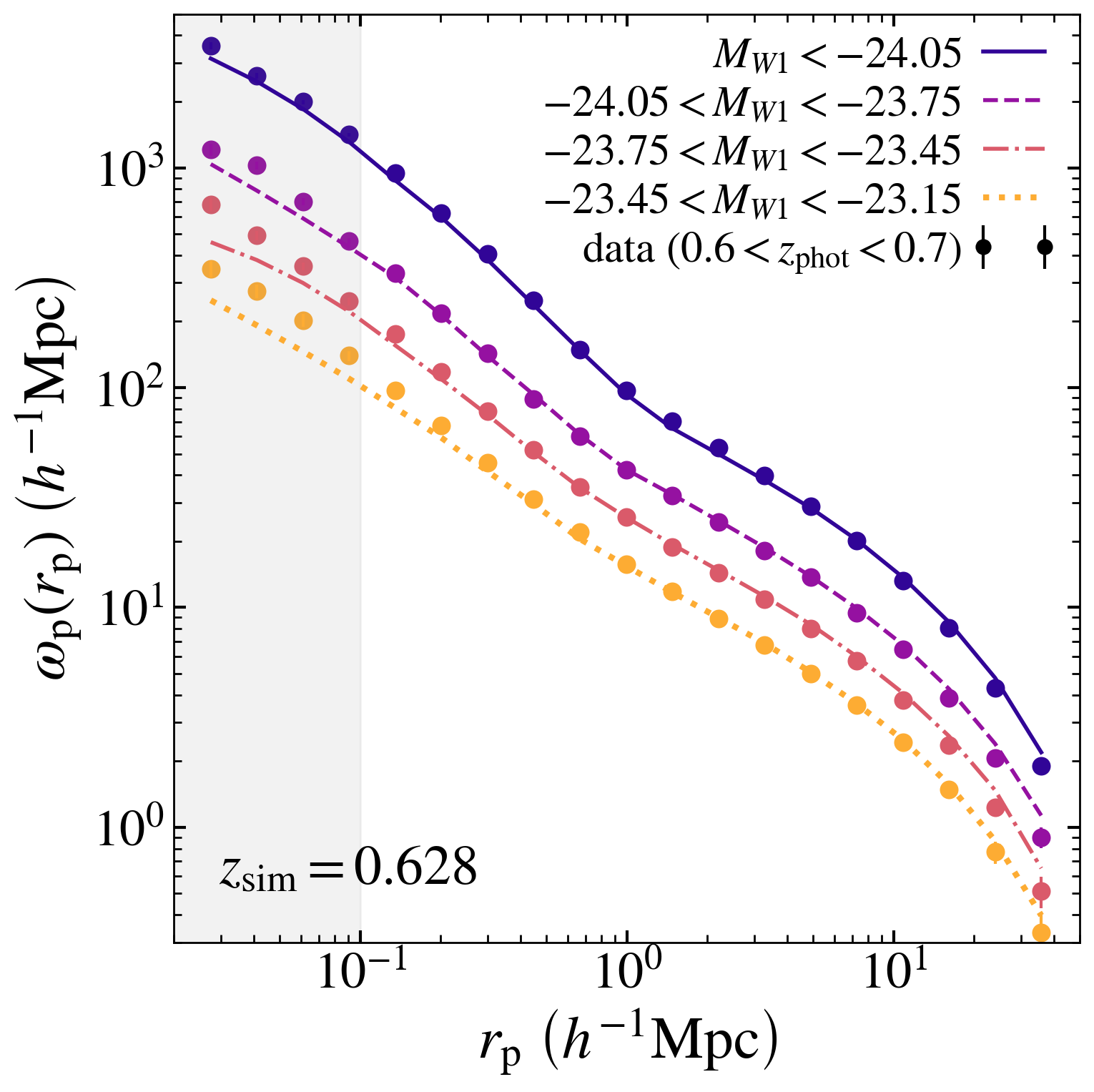}
\caption{Luminosity-binned projected clustering of the $W1$-band absolute magnitude-limited parent galaxy samples and corresponding mock galaxy catalogs, with different colors and line styles representing different luminosity bins as denoted by the legend in each panel. Luminosity bins are offset by 0.15 dex for clarity.
Each panel shows a different redshift bin.
The shaded region at $\rp < 0.1\ \Mpch$ denotes measurements not used for modeling.
}
\label{fig:wp_mag_bins_ir}
\end{figure*}

\begin{figure*}
\centering
  \includegraphics[width=0.3549\linewidth]{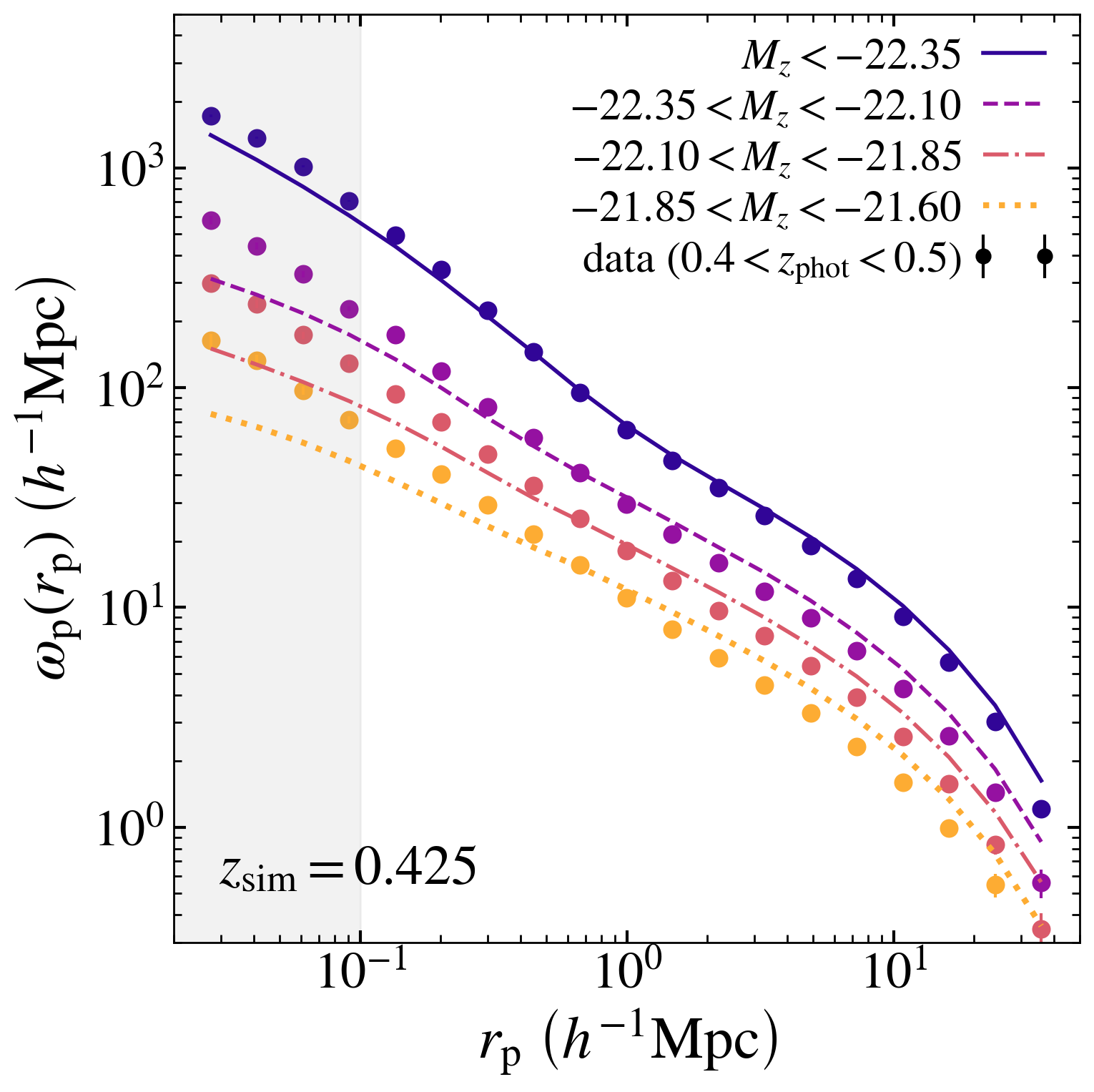}
  \includegraphics[width=0.30\linewidth, trim=3cm 0 0 0, clip]{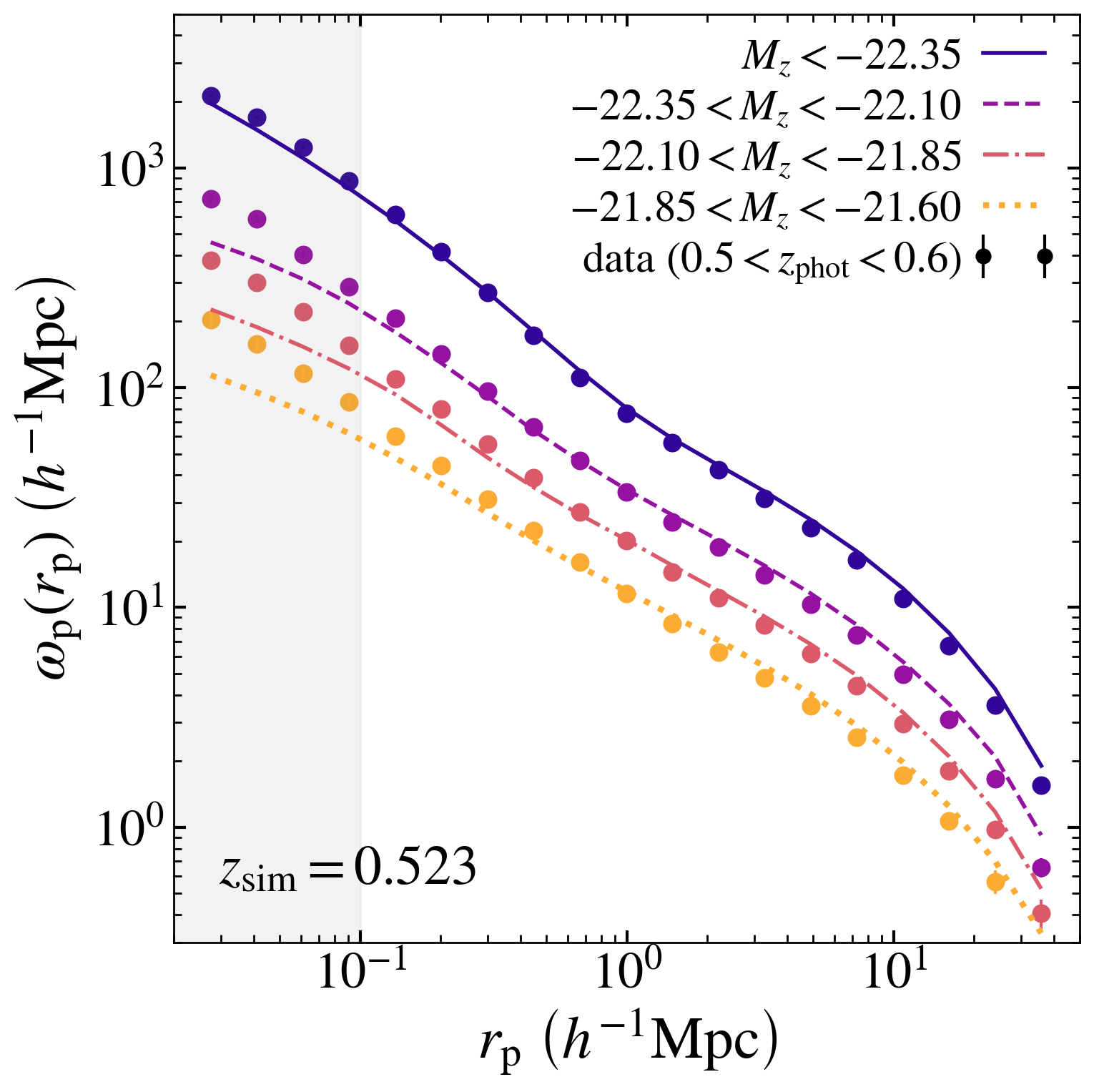}
  \includegraphics[width=0.30\linewidth, trim=3cm 0 0 0, clip]{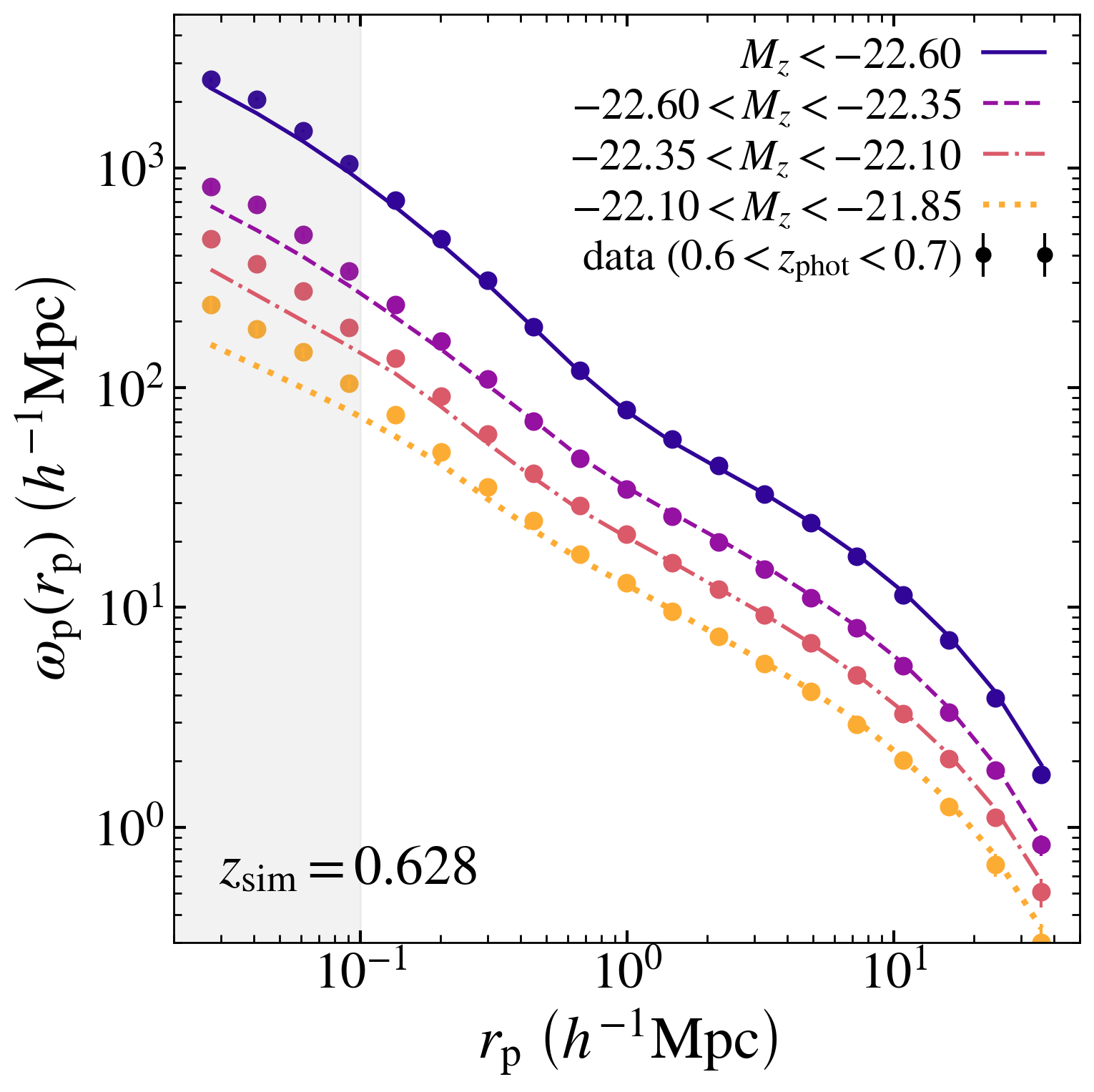}
\caption{Same as Figure~\ref{fig:wp_mag_bins_ir} but for the $z$-band parent galaxy samples and corresponding mock galaxy catalogs.
}
\label{fig:wp_mag_bins_opt}
\end{figure*}

\begin{figure}
  \centering
    \includegraphics[width=\linewidth]{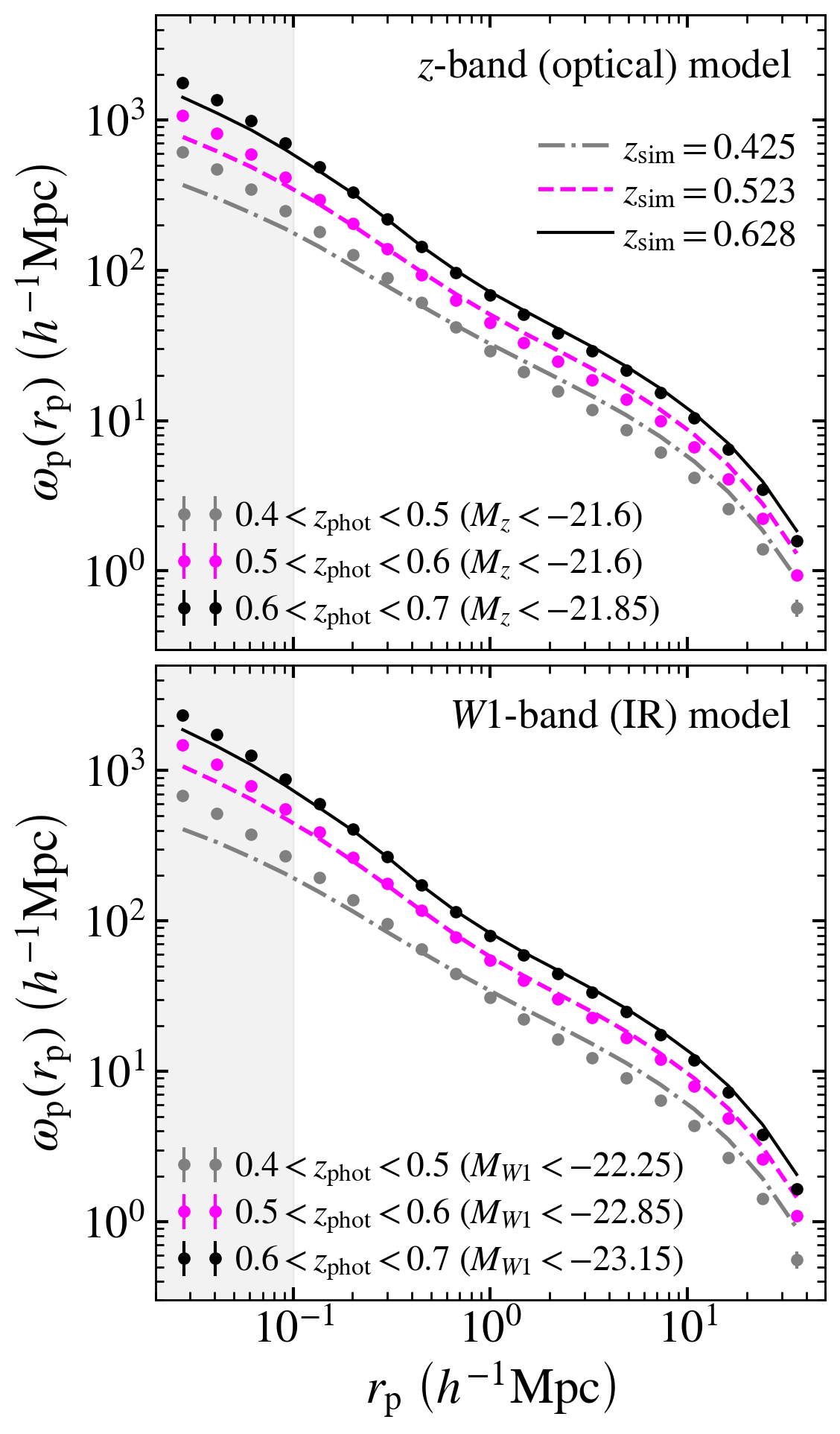}
\caption{Projected clustering of the $z$-band (top panel) and $W1$-band (bottom panel) absolute magnitude-limited parent galaxy samples and corresponding mock galaxy catalogs, with different colors and line styles representing different redshift bins as denoted by the legend in each panel.
Redshift bins are offset by 0.15 dex for clarity.
}
\label{fig:wp_full_sample}
\end{figure}

\begin{deluxetable}{ c c r r r }[h]
\tablecaption{Best-fit values for SHAM model parameters.
\label{tab:scatter_params}
}
\tablehead{
\colhead{Redshift bin} & \colhead{Model} & \colhead{\sigmamag} & \colhead{$s_{\rm LOS}$} & \colhead{$\sigma_{{\rm LOS},0}$}
}
\startdata
\multirow{2}{*}{$0.4 < \zphot < 0.5$} & $W1$-band & 0.66 & -22.3 & -439.2 \\
& $z$-band & 0.57 & -11.4 & -181.5 \\
\vspace{-1ex} \\
\hline
\vspace{-1ex} \\
\multirow{2}{*}{$0.5 < \zphot < 0.6$} & $W1$-band & 0.76 & 21.3 & 584.0 \\
& $z$-band & 0.76 & 60.9 & 1435.6 \\
\vspace{-1ex} \\
\hline
\vspace{-1ex} \\
\multirow{2}{*}{$0.6 < \zphot < 0.7$} & $W1$-band & 0.80 & 31.2 & 839.4 \\
& $z$-band & 0.77 & 50.8 & 1248.4 \\
\enddata
\end{deluxetable}

We use SHAM to create mock galaxy populations with the same number density and luminosity distribution of the parent galaxy sample by assuming the following relation:
\begin{equation}\label{eq:sham}
\neff(<M)=n_{\rm h}(>\vpeak),
\end{equation}
\noindent i.e., the (effective) number density of galaxies, \neff, of magnitude $M$ or brighter equals the number density of halos, $n_{\rm h}$, with circular velocity \vpeak or greater.

To assign absolute magnitudes to halos in each redshift bin we implement the following procedure:
\begin{enumerate}
\item \label{step:neff}
Compute for each parent galaxy sample the effective galaxy number density as a function of absolute magnitude in band ${X \in \{z, W1\}}$, $\neff(<M_X)$. We compute $\neff(<M_X)$ for each parent galaxy sample as follows. For each galaxy in the sample we calculate an effective volume, $V_{\rm eff}$:
\begin{equation}
{V_{\rm eff}(M_X) = f_\Omega \left( V_{\rm max}(M_X)-V_{\rm min} \right)},
\end{equation}
where $f_\Omega$ is the fractional solid angle covered by the parent galaxy sample, and $V_{\rm min}$ is the comoving volume of the lower limit of the redshift bin. $V_{\rm max}(M_X)$ is the comoving volume of either the upper limit of the redshift bin, or of the maximum possible redshift a galaxy of magnitude $M_X$ could have and still be observed at the magnitude limit of the sample, whichever is smaller.
The effective galaxy number density is the sum of the inverse of $V_{\rm eff}(M_X)$ over all galaxies in the sample:
\begin{equation}\label{eq:neff}
  \neff(<M_X) = \sum_i \left[ V^i_{\rm eff}(M_X) \right]^{-1}.
\end{equation}
\item \label{step:nh}
Assign absolute magnitudes to mock galaxies with no scatter in the luminosity--\vpeak relation according to Eq.\ \ref{eq:sham}. For each halo we find the cumulative number density $n_{\rm h}(>\vpeak)$ corresponding to its value of \vpeak. We then assign to each halo a mock galaxy with the absolute magnitude $M_X$ at which the effective number density $\neff(<M_X)$ of the parent galaxy sample equals $n_{\rm h}(>\vpeak)$.
\item \label{step:sham_scatter}
To incorporate magnitude-dependent scatter into the luminosity--\vpeak relation, we assign to each mock galaxy a new absolute magnitude, $M'_X$, where $M'_X$ is drawn from a Gaussian distribution centered at $M_X$ with width \sigmamag. \sigmamag is proportional to the absolute value of $M_X$, and the constant of proportionality is a free parameter in the model.
We then rank order all mock galaxies (including their \vpeak values) by $M'_X$, rank order the original distribution of mock magnitudes $M_X$, and assign the ordered original mock magnitude distribution to the mock galaxy catalog ordered by $M'_X$. This method incorporates scatter into the luminosity--\vpeak relation while replicating the target luminosity function exactly in the resulting mock galaxy catalog.
\item \label{step:los_scatter}
Scatter the positions of mock galaxies along one of the three axes of the simulation volume to mimic the uncertainty in radial (line-of-sight) position of our target galaxy samples due to photometric redshift errors.
For each mock galaxy we draw a ``scattered" coordinate $x'$ from a Gaussian distribution of width \sigmalos centered at the galaxy's original $x$ position.
Mock galaxies that scatter out of the simulation volume of $1~h^{-3}\ {\rm Gpc}^3$ are wrapped back in to preserve the periodic boundary conditions, e.g., a mock galaxy at $x=25~\Mpch$ that scatters to $x'=-50~\Mpch$ is placed at $x=950~\Mpch$.
We repeat this process for the other two simulation axes (i.e., $y$ and $z$).
\item \label{step:model_wp}
For each mock catalog, compute the projected correlation function, \wprp (see \S\ref{subsec:wprp}), averaged over the three axes of the simulation volume as described in the previous step.
We then compute the goodness-of-fit per degree of freedom, \chisqred (see \S\ref{subsec:error}), of the model fit to the mean \wprp of the relevant parent galaxy sample. We measure \chisqred for each mean \wprp in bins of absolute magnitude ($M_z$ or $M_{W1}$).
\item \label{step:model_refine}
Repeat steps \ref{step:sham_scatter} through \ref{step:model_wp} for additional values of \sigmamag and \sigmalos as needed to minimize \chisqred in each luminosity bin. In practice we first coarsely sample a wide range of values of \sigmamag and \sigmalos:
\begin{subequations}
  \begin{align}
    0 \leq \frac{\sigmamag}{|M_X|} \leq 1.0,\ & \Delta \!\! \left[\frac{\sigmamag}{|M_X|}\right]=0.1 \\
    0 \leq \sigmalos \leq 150~\Mpch,\ & \Delta \sigmalos=10~\Mpch.
  \end{align}
\end{subequations}
We then more densely sample (using $\Delta(\sigmamag/|M_X|)=0.01$ and ${\Delta \sigmalos=5~\Mpch}$) narrower ranges of both parameters around the initial coarse-grained values that minimize \chisqred.
\item \label{step:sigma_linear}
Finally, identify the value of \sigmamag that coincides with the region of \sigmamag--\sigmalos parameter space containing the minimum \chisqred across all luminosity bins, and parameterize the dependence of \sigmalos on absolute magnitude, $M_X$, as follows:
\begin{equation}\label{eq:sigmalos}
  \sigmalos(M_X) = s_{\rm LOS}\, M_X + \sigma_{{\rm LOS},0}.
\end{equation}
\noindent The best-fit values of \sigmamag, $s_{\rm LOS}$, and $\sigma_{{\rm LOS},0}$ are given in Table~\ref{tab:scatter_params}. The values of $s_{\rm LOS}$ and $\sigma_{{\rm LOS},0}$ are determined by linear fits to $\sigmalos^i$ versus $\langle M^i_X \rangle$, where $\sigmalos^i$ are the values that minimize \chisqred of \wprp at fixed \sigmamag\footnote{We tested using \sigmamag with linear dependence on absolute magnitude instead of a constant value across all luminosity bins, and found that fixing \sigmamag while allowing \sigmalos to scale linearly with luminosity reproduces \wprp (\S\ref{subsec:wprp}) better than if both parameters have linear luminosity dependence.} in the $i$th luminosity bin, $M^i_X$.
\end{enumerate}

The luminosity assignment stage of our modeling procedure involves three free parameters, \sigmamag, $s_{\rm LOS}$, and $\sigma_{{\rm LOS},0}$, which account for scatter in the luminosity--\vcirc relation and photometric redshift errors of our target galaxy samples. We constrain these parameters by fitting the projected correlation functions of mock galaxy catalogs created from our model to those of the corresponding parent galaxy samples.

Figures~\ref{fig:wp_mag_bins_ir} and \ref{fig:wp_mag_bins_opt} show \wprp for the $W1$- and $z$-band absolute magnitude-limited parent galaxy samples, respectively, and corresponding mock galaxy catalogs. The agreement between the data and model increases with increasing luminosity within each redshift bin, and with increasing redshift overall.
Note that the shaded regions in Figures~\ref{fig:wp_mag_bins_ir} and \ref{fig:wp_mag_bins_opt} at $\rp < 0.1\ \Mpch$ denotes measurements not used for model fitting.

The clustering of the full magnitude-limited parent galaxy samples and corresponding mock catalogs is shown for each redshift bin in Figure~\ref{fig:wp_full_sample}, with \wprp for each redshift bin offset by 0.15 dex for clarity. Agreement between the model and data increases with increasing redshift. We emphasize that while the model \wprp deviates from that of the data for the full magnitude-limited parent samples in the ${0.4 < \zphot < 0.5}$ redshift bin, there is still good data--model agreement within the highest luminosity bins, where the vast majority of LRGs reside, for both the $W1$- and $z$-band models (Figures~\ref{fig:wp_mag_bins_ir} and \ref{fig:wp_mag_bins_opt}, respectively).

\subsection{Color assignment}\label{subsec:color_assign}

\begin{figure}
\centering
\includegraphics[width=\linewidth]{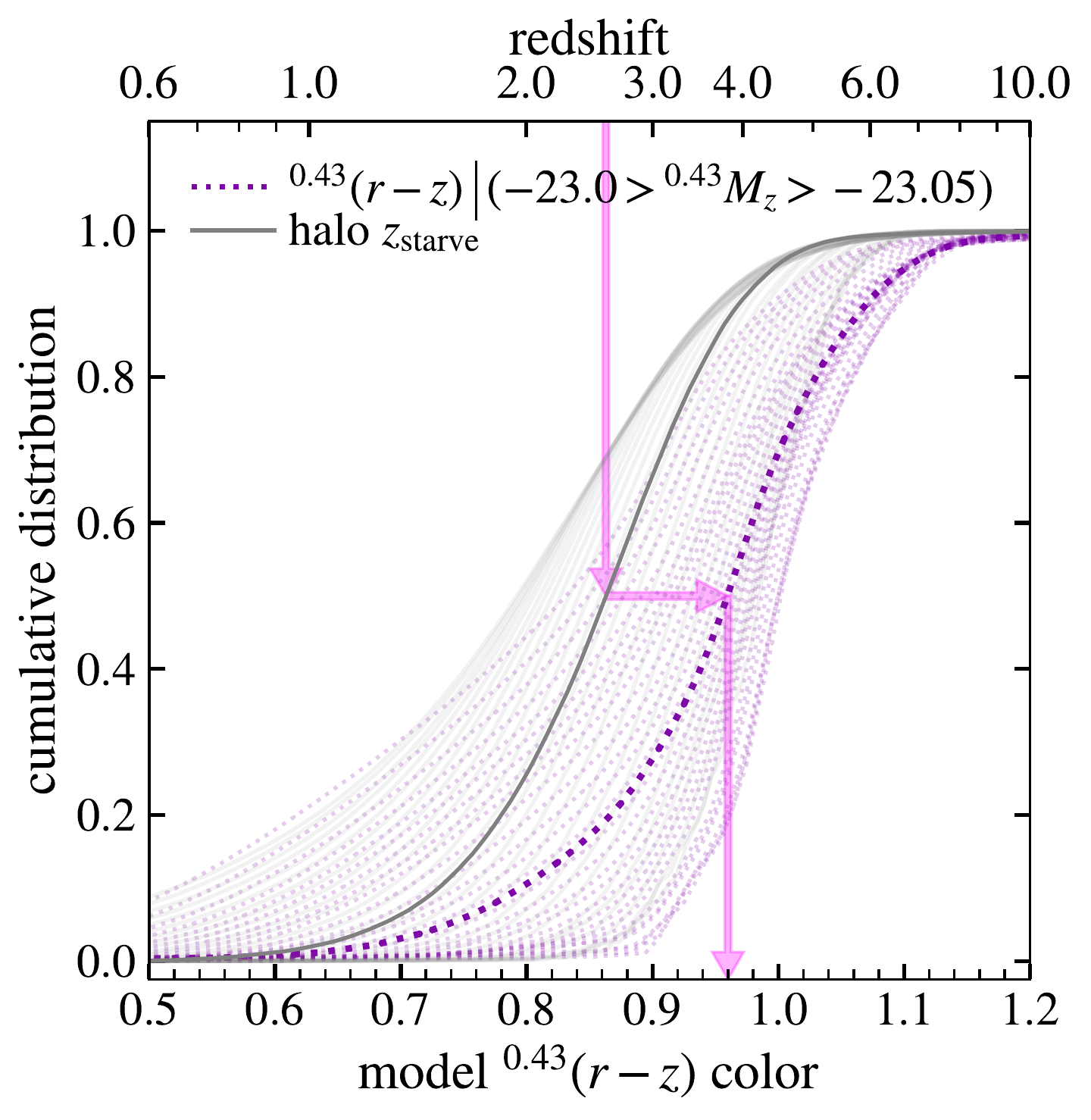}
\caption{
Illustration of the model color assignment procedure (\S\ref{subsec:color_assign}), which equates halo \zstarve with galaxy color at fixed luminosity. The purple dotted curve is the cumulative distribution of $^{0.43}(r-z)$ color for galaxies in the $-23.0 > {^{0.43}M_z} > -23.05$ luminosity bin, while the gray solid curve is the cumulative distribution of halo \zstarve for mock galaxies in the same luminosity bin. The magenta arrows indicate that a mock galaxy in this luminosity bin in a halo with $\zstarve\approx2.6$ is assigned a $^{0.43}(r-z)$ color of $0.96$. The faint purple and gray curves are the galaxy color and halo \zstarve distributions of additional luminosity bins.
}
\label{fig:color_assign}
\end{figure}

Figure~\ref{fig:color_assign} shows an illustration of our color assignment algorithm (Eq.\ \ref{eq:color_assign}) with age distribution matching for the ${0.4<\zphot<0.5}$ redshift bin of the $z$-band model. The dotted purple curve is the cumulative distribution of $^{0.43}(r-z)$ color for galaxies in the $-23.0 > {^{0.43}M_z} > -23.05$ luminosity bin, and the solid gray curve is the halo \zstarve distribution of $\zsim=0.425$ mock galaxies in the same model luminosity bin. The magenta arrows in Figure~\ref{fig:color_assign} indicate that a mock galaxy in this luminosity bin in a halo with $\zstarve\approx2.6$ is assigned a $^{0.43}(r-z)$ color of 0.96.

Implementation of the color assignment algorithm with no scatter in the color--\zstarve relationship yields mock LRG samples (see \S\ref{subsec:mock_lrg_select} below) that generally overpredict the clustering amplitude compared to the data, with the exception optical LRGs selected from the $W1$-band model (see Figure~\ref{fig:wp_lrg}). We tested the effect of introducing scatter into the color--\zstarve relation on the clustering of mock LRGs using an analogous method to how scatter is incorporated into the luminosity--\vpeak relation (see \S\ref{subsec:luminosity_assign}). After assigning model colors $\mathcal{C}$ without scatter according to Eq.\ \ref{eq:color_assign} as described above, we assign to each mock galaxy a new color $\mathcal{C}'$, where $\mathcal{C}'$ is drawn from a Gaussian distribution centered at $\mathcal{C}$ with width $\sigma_{\rm color}$. We then rank order all mock galaxies by $\mathcal{C}'$, rank order the original distribution of mock colors $\mathcal{C}$, and assign the ordered original color distribution to the mock galaxy catalog ordered by $\mathcal{C}'$.

For mock LRGs (both optical and IR) from the $z$-band model, as well as mock IR LRGs from the $W1$-band model, clustering amplitude decreases with increasing $\sigma_{\rm color}$ until $\sigma_{\rm color}$ is sufficiently large that model colors are effectively assigned at random, at which point the decrease in clustering amplitude levels off at a constant value that still overpredicts \wprp relative to the data.
The exception to this is the predicted clustering of mock optical LRGs from the $W1$-band model, which largely agrees with the data with \emph{no} scatter in the color--\zstarve relation.
Additionally, $\sigma_{\rm color} > 0$ in this instance does \emph{not} affect the predicted clustering amplitude of mock optical LRGs.
This implies that color is uncorrelated with halo age for LRGs, which we discuss further in \S\ref{subsec:clust_lrg}, and motivates our decision to proceed with two versions of our models (rather than introduce and constrain $\sigma_{\rm color}$ as an additional parameter):\
(1) a ``default" age distribution model, in which galaxy color increases monotonically with increasing halo \zstarve at fixed luminosity (i.e., $\sigma_{\rm color} = 0$), and (2) a ``random color'' model, in which there is no correlation between galaxy color and halo \zstarve.
Both models yield magnitude-limited mock galaxy catalogs with identical color--magnitude distributions that reproduce the target distribution of the relevant data; the only difference is whether galaxy color is maximally correlated (default model) or entirely uncorrelated (random color model) with halo \zstarve.

\subsection{Selecting mock LRGs}\label{subsec:mock_lrg_select}

\begin{figure*}
\centering
  \setlength{\tabcolsep}{0pt}
    \begin{tabular}{ c c }
      {\large $z$-band model} & {\large $W1$-band model} \\
      \includegraphics[width=0.49\linewidth, trim=0 2.578cm 0 0, clip]{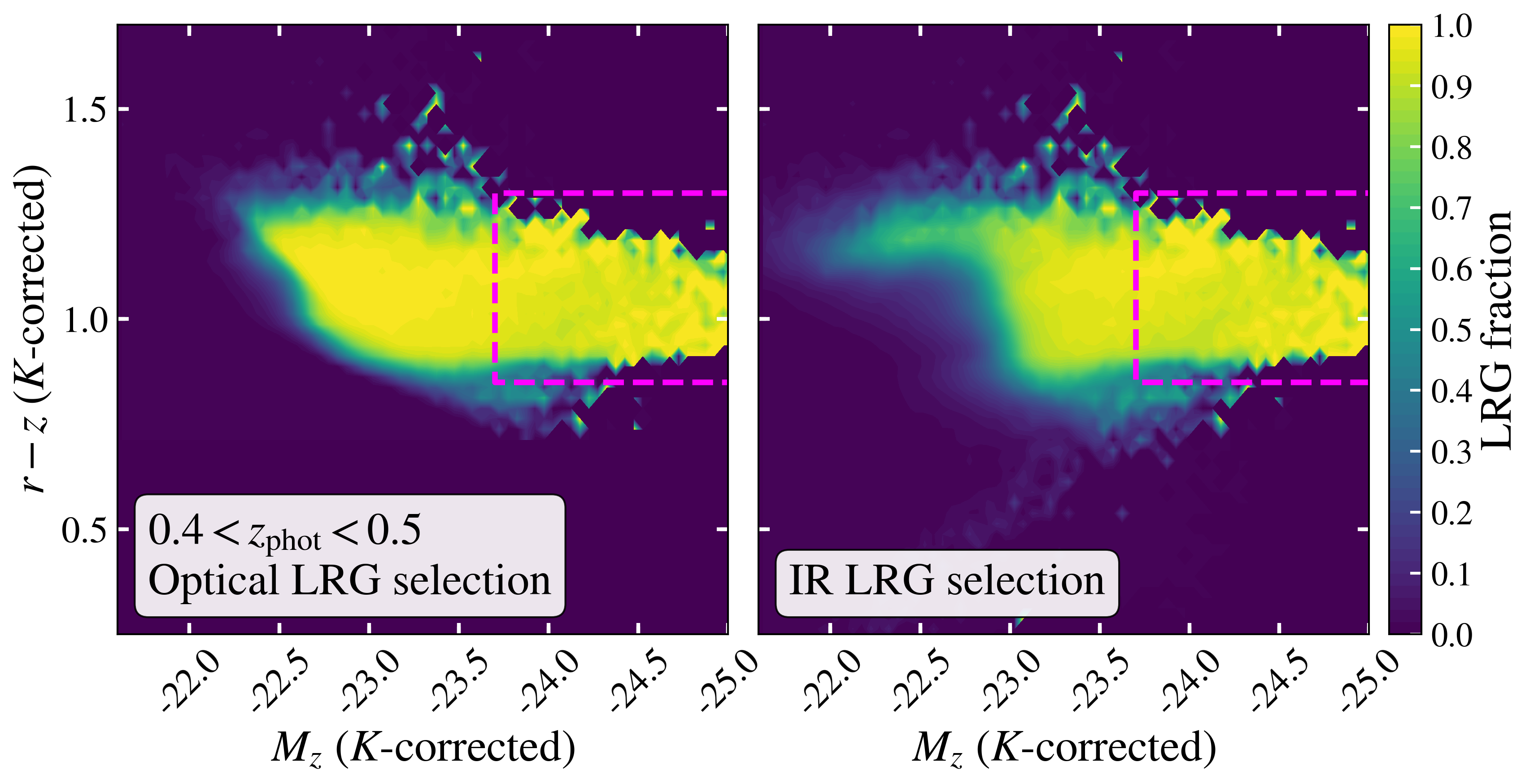} &
      \includegraphics[width=0.49\linewidth, trim=0 2.578cm 0 0.2mm, clip]{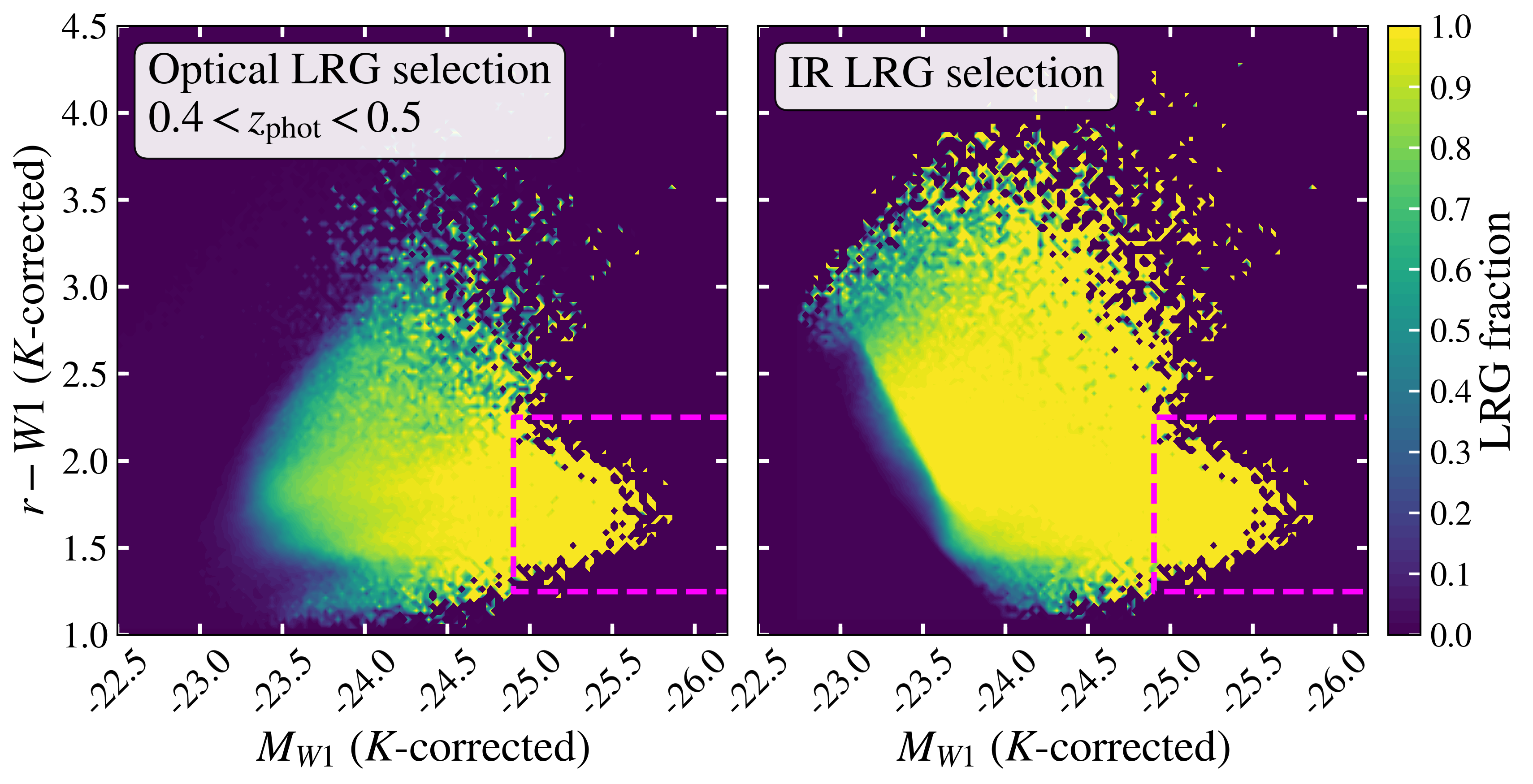} \\
      \includegraphics[width=0.49\linewidth, trim=0 2.578cm 0 0, clip]{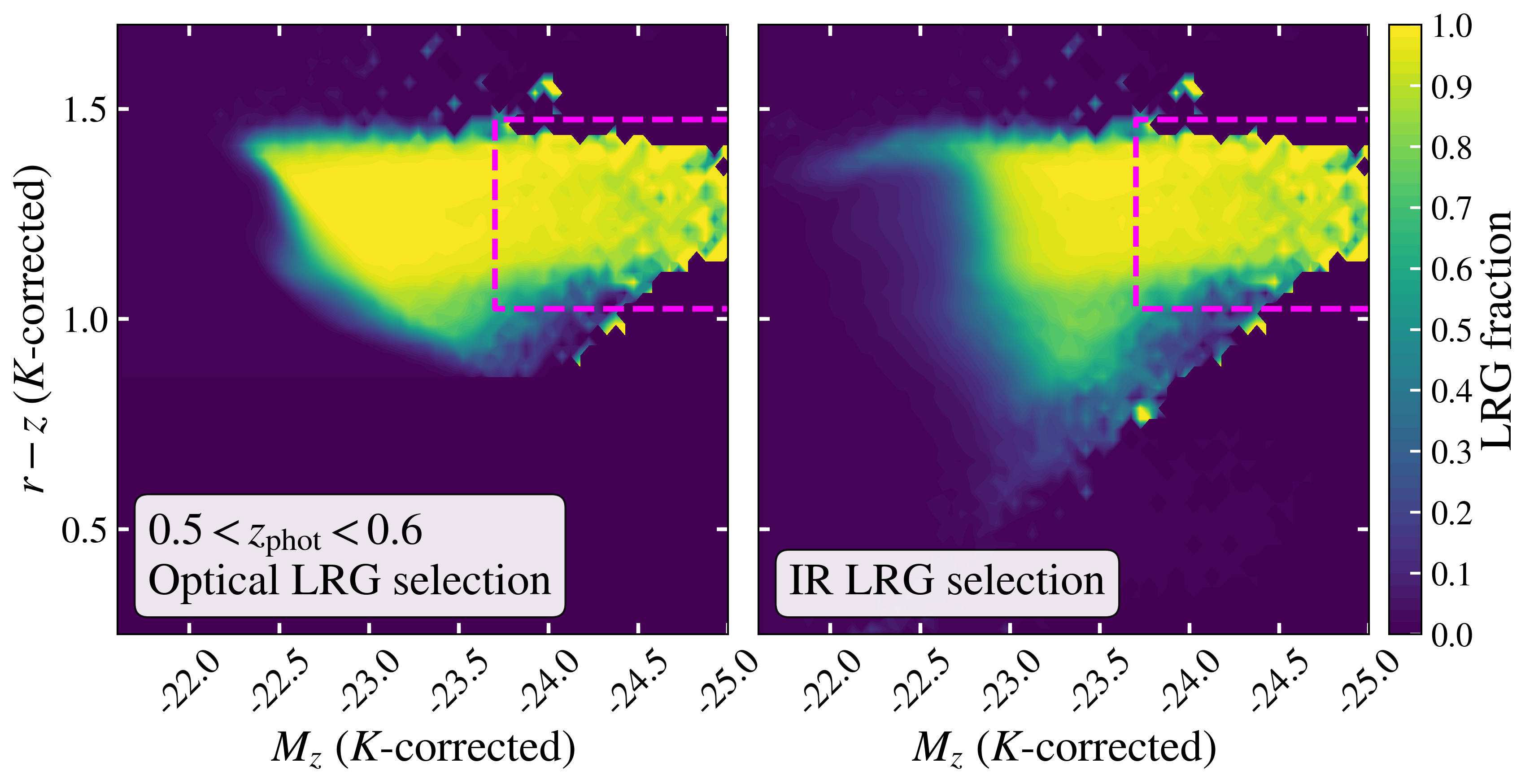} &
      \includegraphics[width=0.49\linewidth, trim=0 2.578cm 0 0.2mm, clip]{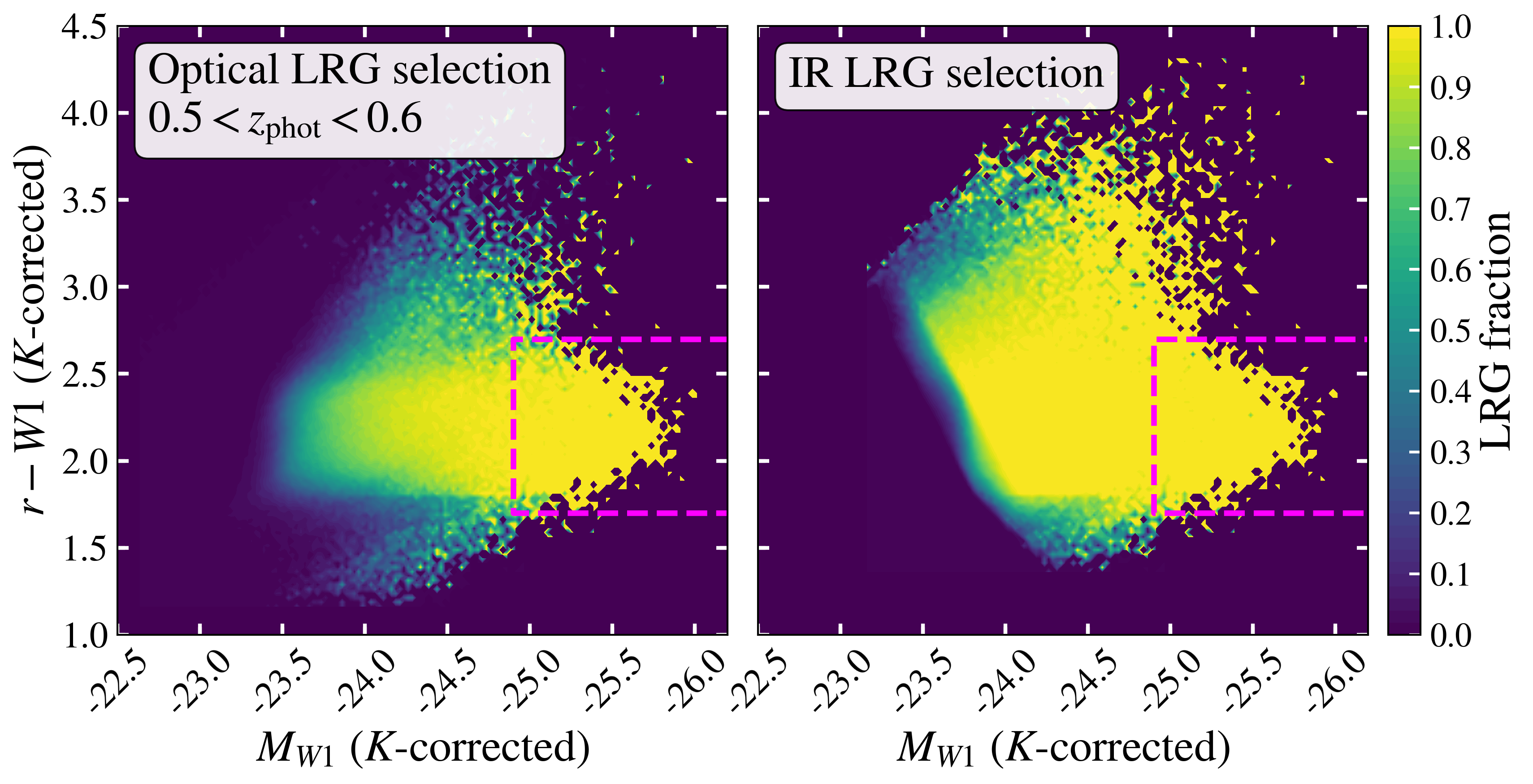} \\
      \includegraphics[width=0.49\linewidth]{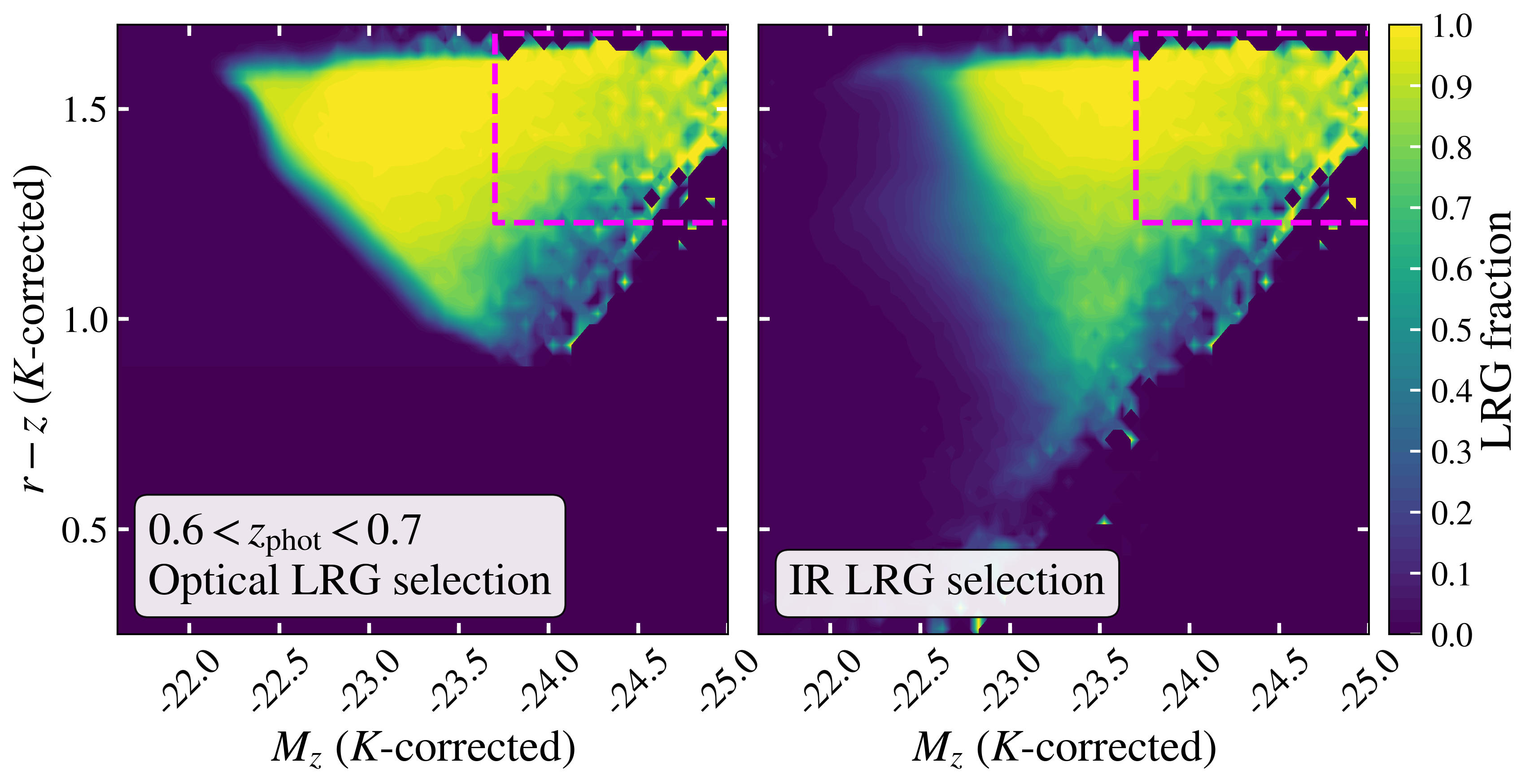} &
      \includegraphics[width=0.49\linewidth]{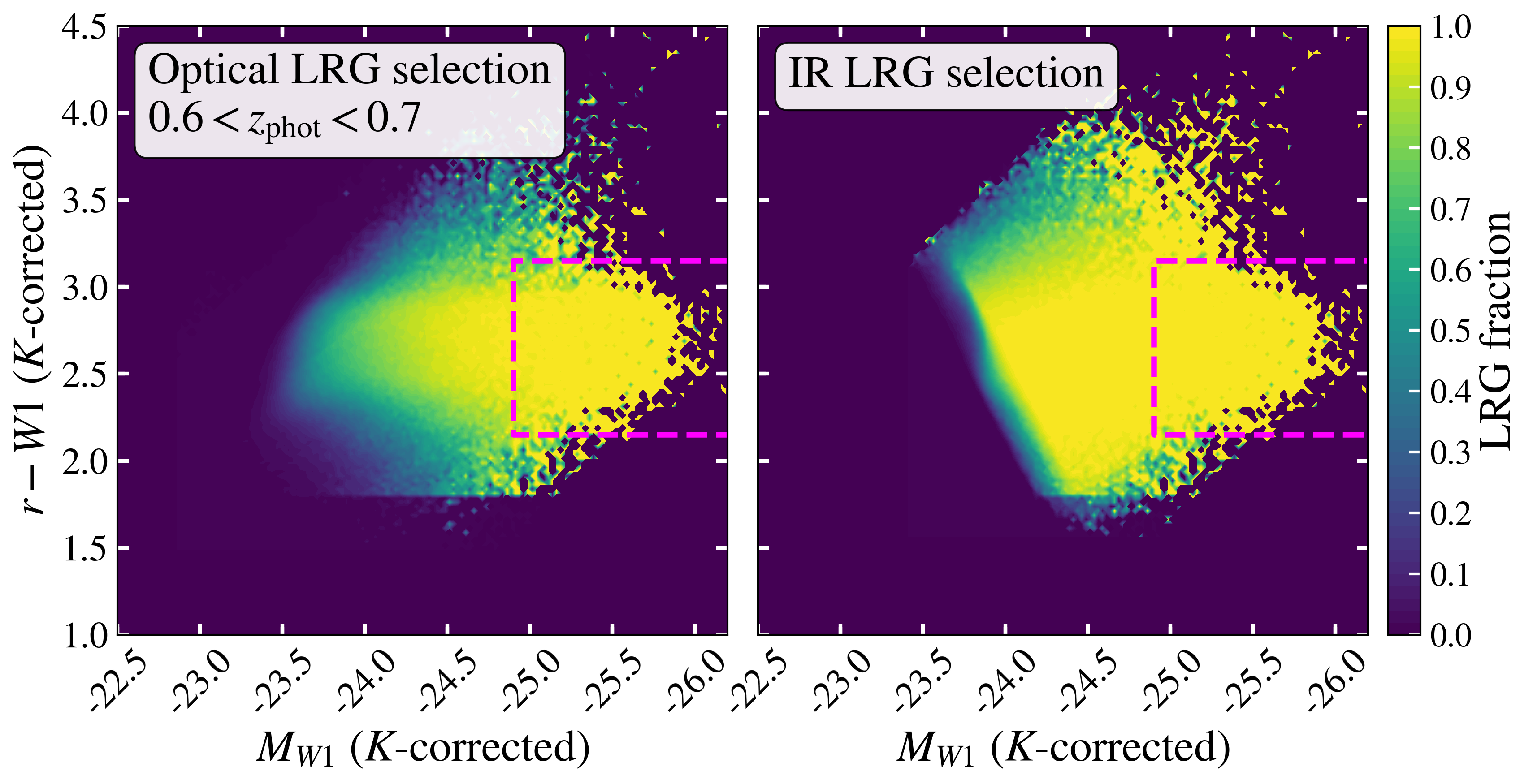}
    \end{tabular}
\caption{
LRG fractions in color--magnitude space used to select optical and IR LRG samples from our magnitude-limited mock galaxy catalogs.
The first and second columns show optical and IR LRG targets, respectively, in ${r-z}$ color versus $M_z$ magnitude space.
The third and fourth columns show optical and IR LRG targets, respectively, in ${r-W1}$ color versus $M_{W1}$ magnitude space.
Each row shows a different redshift bin.
Note that the distributions of optical LRGs in the $z$-band model (first column) and IR LRGs in the $W1$-band model (fourth column) are compact by design, i.e., the fraction of galaxies that are LRGs is either $\sim0$ or $\sim1$ nearly everywhere in color--magnitude space. In contrast, the distributions of IR LRGs in the $z$-band model (second column) and optical LRGs in the $W1$-band model (third column) have a broader gradient between 0 and 1, especially toward fainter magnitudes.
The boxed region in each panel marks the luminous end of the red sequence, where essentially all galaxies are expected to be LRGs. The LRG fraction in this boxed region is at least 98\% for the $W1$-band model (third and fourth columns), but only {92--94\%} for the $z$-band model (first and second columns).}
\label{fig:lrg_frac}
\end{figure*}

\begin{figure*}
\centering
\includegraphics[width=\linewidth]{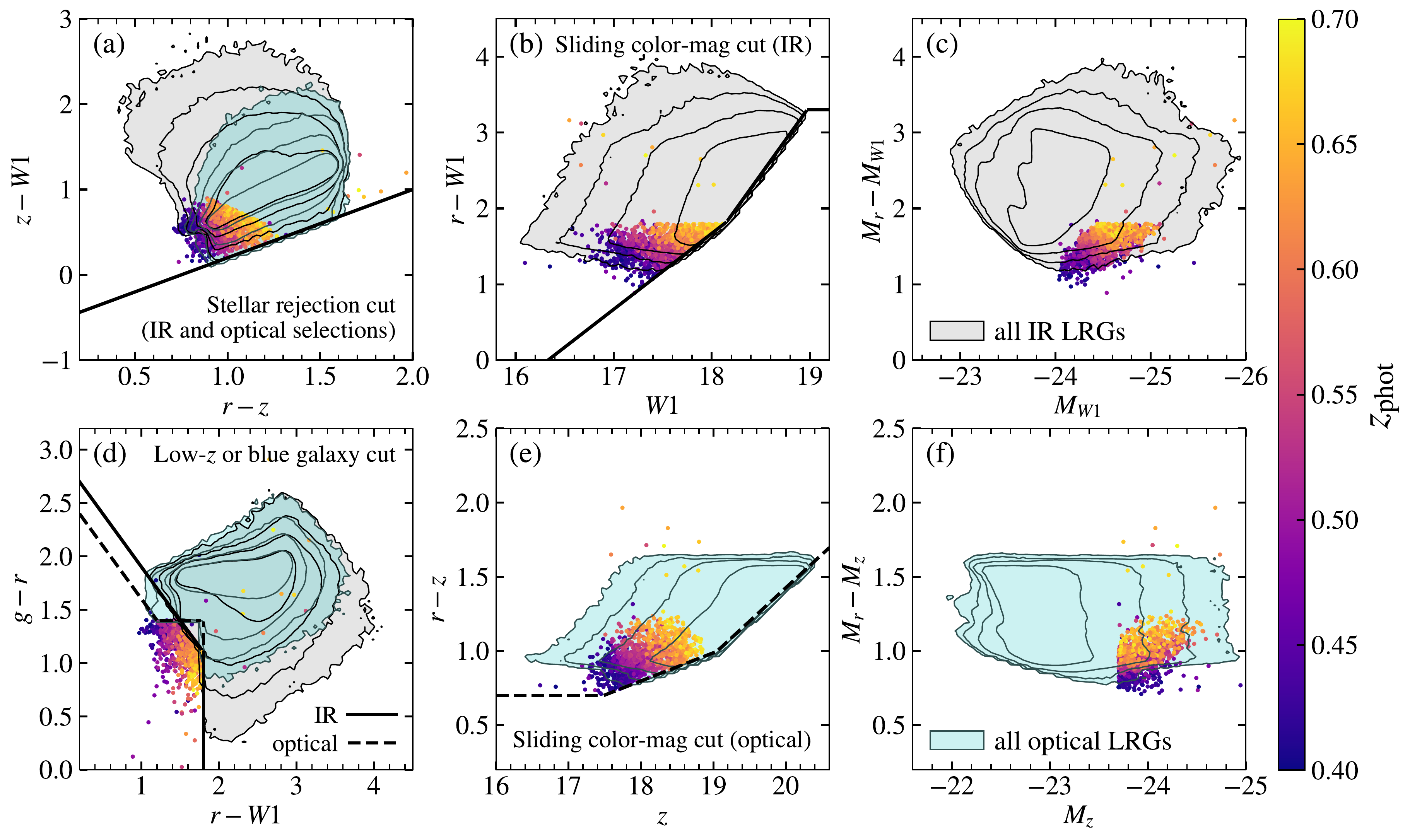}
\caption{
The color-coded points in each panel are bright (${M_z<-23.7}$), red-sequence objects with photometric redshifts between 0.4 and 0.7 that are \emph{not} DESI LRG targets, i.e., objects with ${M_z<-23.7}$ in the first two columns of Figure~\ref{fig:lrg_frac} that cause the LRG fraction at the bright end of the red sequence to be $<1$.
The dashed and solid black lines in panels (a), (b), (d), and (e) are the various color--color and color--magnitude cuts that define the optical and IR DESI LRG target samples (Eqs.~\ref{eq:lrg_opt} and \ref{eq:lrg_ir}, respectively).
The light blue and gray contours show the distributions of \emph{all} optical and IR DESI LRG targets with photometric redshifts between 0.4 and 0.7.
}
\label{fig:not-lrg_outliers}
\end{figure*}

DESI LRG target selection (see \S\ref{subsec:lrg_target_select} above) is entirely a function of apparent $g$, $r$, $z$, and $W1$ (and $z_{\rm fiber}$) magnitudes. As such it would be ideal to have mock galaxy catalogs with each of these model magnitudes for every galaxy; mock LRG samples could then be obtained simply by applying the DESI LRG target selection functions (Eqs.\ \ref{eq:lrg_opt} and \ref{eq:lrg_ir}) to the full mock for each redshift bin. However the SHAM and age distribution matching techniques we employ create mock galaxy catalogs with absolute $z$- or $W1$-band magnitudes and corresponding ${r-z}$ or ${r-W1}$ colors.

To select mock DESI LRG targets using only these model quantities we first identify where actual DESI LRG targets reside in $K$-corrected color--magnitude space in each redshift bin for both optical and IR DESI LRG targets. In each redshift bin we compute the fractions of galaxies that are optical and IR DESI LRG targets in narrow two-dimensional bins in color--magnitude space. This is shown in Figure~\ref{fig:lrg_frac}.

We then use these optical and IR LRG fractions in color--magnitude space to statistically select mock optical and IR LRGs from our mock galaxy catalogs in the corresponding model color--magnitude space, i.e., in each color--magnitude bin we draw mock galaxies and flag them as LRGs until the LRG fraction in that bin matches the value from the data.

It is worth noting that the LRG fraction in most of color--magnitude space is either $\sim0$ or $\sim1$ by design, especially for optical LRGs in ${r-z}$ versus $M_z$ (i.e., optical) space, and IR LRGs in ${r-W1}$ versus $M_{\rm W1}$ (i.e., IR) space. This is clearly shown in Figure~\ref{fig:lrg_frac}.

The first and second columns of Figure~\ref{fig:lrg_frac} show the fractions of optical and IR LRG targets, respectively, in optical space, i.e., ${r-z}$ color versus $M_z$ magnitude, and the red sequence is clearly visible in each panel. For example, in the ${0.4<\zphot<0.5}$ redshift bin the red sequence corresponds almost entirely to the region where the LRG fraction is $\sim1$, i.e., where ${M_z \lesssim -22.5}$ and ${r-z}$ is between $\sim0.9$ and $\sim1.3$ (with a shift toward redder colors in higher redshift bins).

Interestingly, the LRG fraction begins to deviate from $\sim1$ in the \emph{most luminous} region of optical space ($M_z\lesssim-23.7$), and this is true for both optical and IR LRG targets.
Both selections (Eqs.\ \ref{eq:lrg_opt} and \ref{eq:lrg_ir}) exclude some of the most luminous red-sequence objects in optical space, but \emph{not} in IR space (third and fourth columns of Fig.~\ref{fig:lrg_frac}).

To investigate why these luminous, red-sequence objects in the first two columns of Figure~\ref{fig:lrg_frac}) are not selected as DESI LRG targets we looked at their positions in color--color and color--magnitude space relative to the full set of DESI LRG target selection cuts (Eqs.\ \ref{eq:lrg_opt} and \ref{eq:lrg_ir}).
This is shown in Figure~\ref{fig:not-lrg_outliers}, where light blue and gray contours in each panel show the distributions of \emph{all} optical and IR DESI LRG targets with photometric redshifts between 0.4 and 0.7.

The color-coded points in each panel are luminous (${M_z<-23.7}$), red-sequence objects in the same redshift range that are \emph{not} DESI LRG targets despite passing the stellar cut (Eqs.\ \ref{eq:lrg_opt_stellar} and \ref{eq:lrg_ir_stellar}; solid black line in panel (a))
and the optical and IR color--magnitude cuts (Eq.\ \ref{eq:lrg_opt_color-mag}; dashed black line in panel (e), and Eq.\ \ref{eq:lrg_ir_color-mag}; solid black line in panel (b)),
as well as the $z_{\rm fiber}$ cut (Eqs.\ \ref{eq:lrg_opt_fiber} and \ref{eq:lrg_ir_fiber}; not shown).

Panel (d) of Figure~\ref{fig:not-lrg_outliers} shows that the low-redshift or blue color cut (Eqs.\ \ref{eq:lrg_opt_lowz} and \ref{eq:lrg_ir_lowz}) excludes these objects from selection as DESI LRG targets.

\section{Predicted LRG properties}\label{sec:predict}
In this section we present the results of our $z$-band and $W1$-band models. We report the HOD parameters of mock LRGs and compare the predicted clustering to the data and to relevant studies in the literature.

\subsection{LRG clustering}\label{subsec:clust_lrg}

\begin{figure*}
  \centering
    \includegraphics[width=\linewidth]{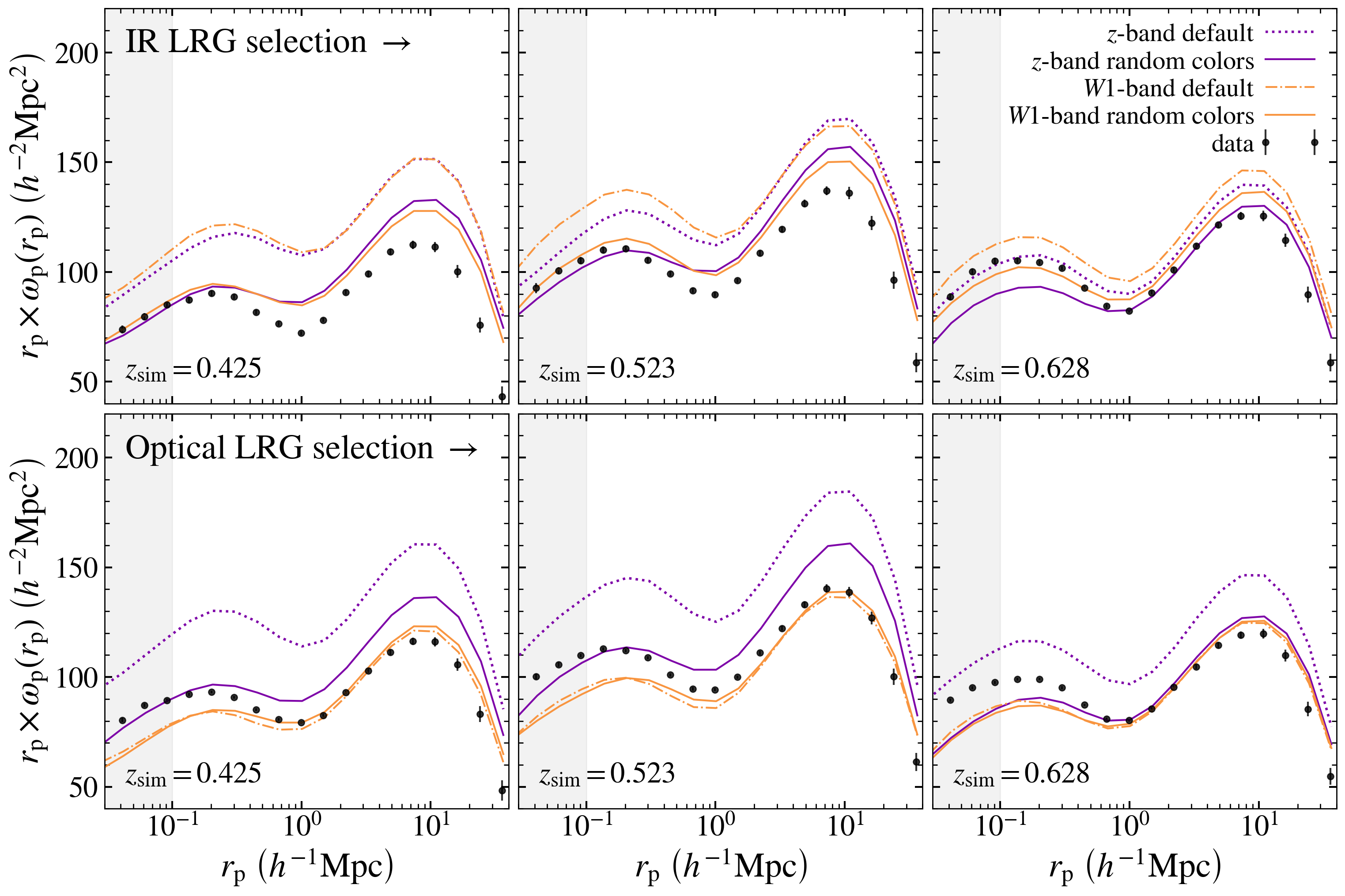}
    \caption{Predicted clustering of IR (top row) and optical (bottom row) mock LRGs compared to the clustering of the relevant LRG target samples from the data (black points in each panel).
    Dotted purple lines in each panel show mock LRGs from the default $z$-band model (monotonic correspondence between $r-z$ color and \zstarve at fixed luminosity), while solid purple lines show LRGs from the $z$-band model with randomized colors.
    Dashed orange lines in each panel show mock LRGs from the default $W1$-band model, and solid orange lines show LRGs from the $W1$-band model with randomized colors.
    Each column shows a different redshift bin.
    }
\label{fig:wp_lrg}
\end{figure*}

\begin{figure}
\centering
\includegraphics[width=\linewidth]{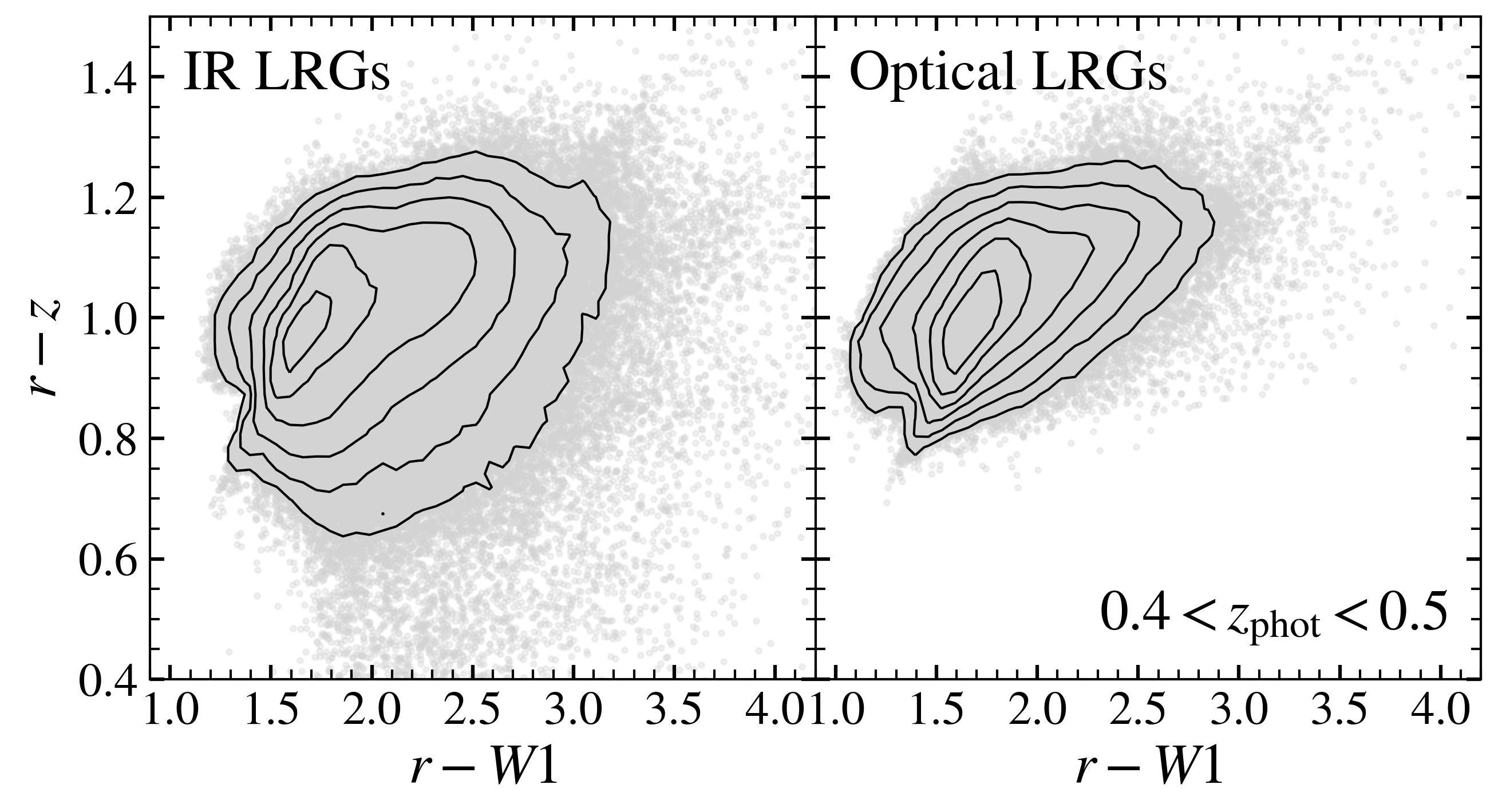}
\caption{IR color ($r-W1$) versus optical color ($r-z$) for IR (left panel) and optical (right panel) LRGs at $0.4<\zphot<0.5$.
}
\label{fig:color_color}
\end{figure}

\begin{figure}
  \centering
    \includegraphics[width=\linewidth]{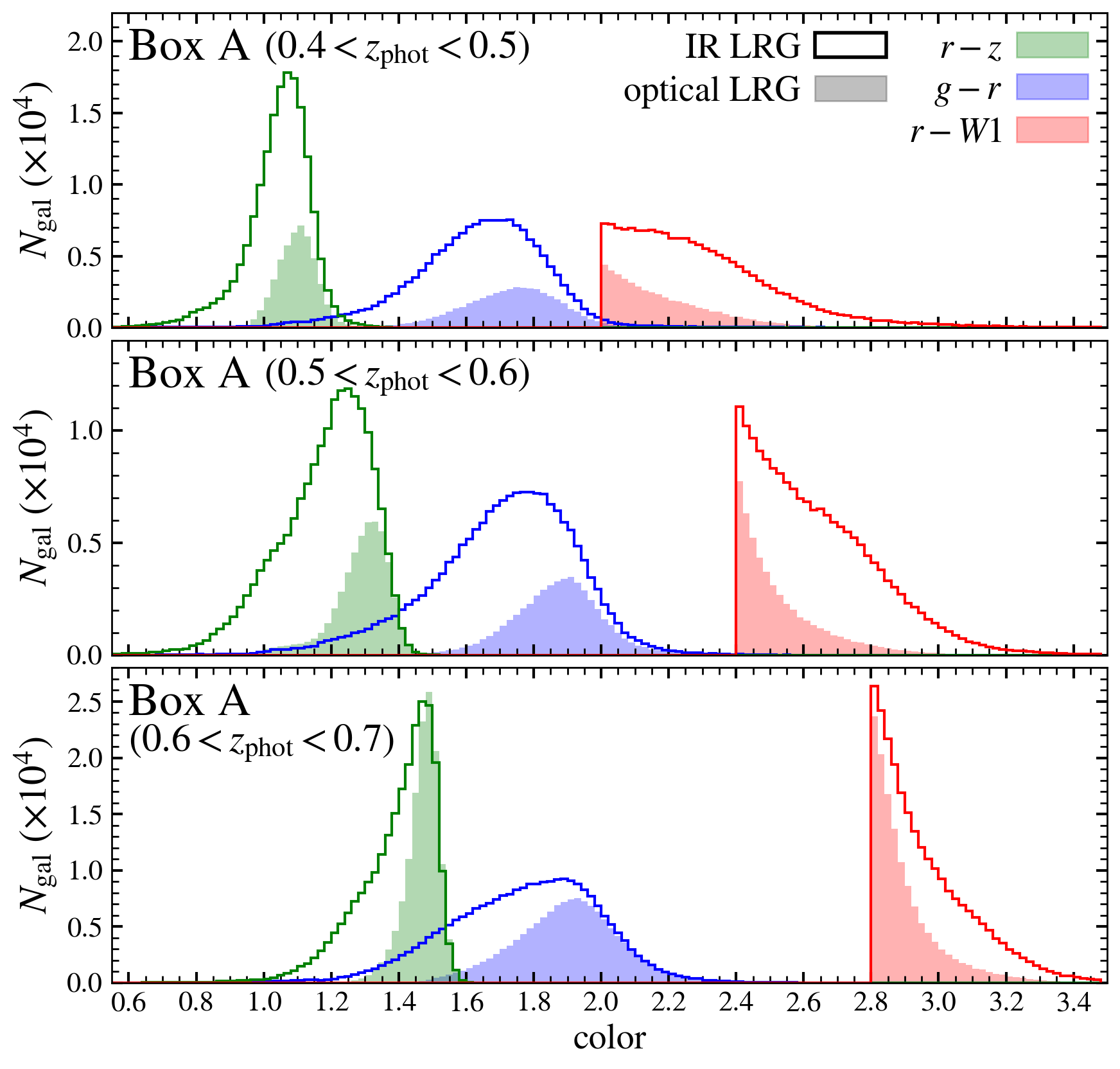}
    \includegraphics[width=\linewidth]{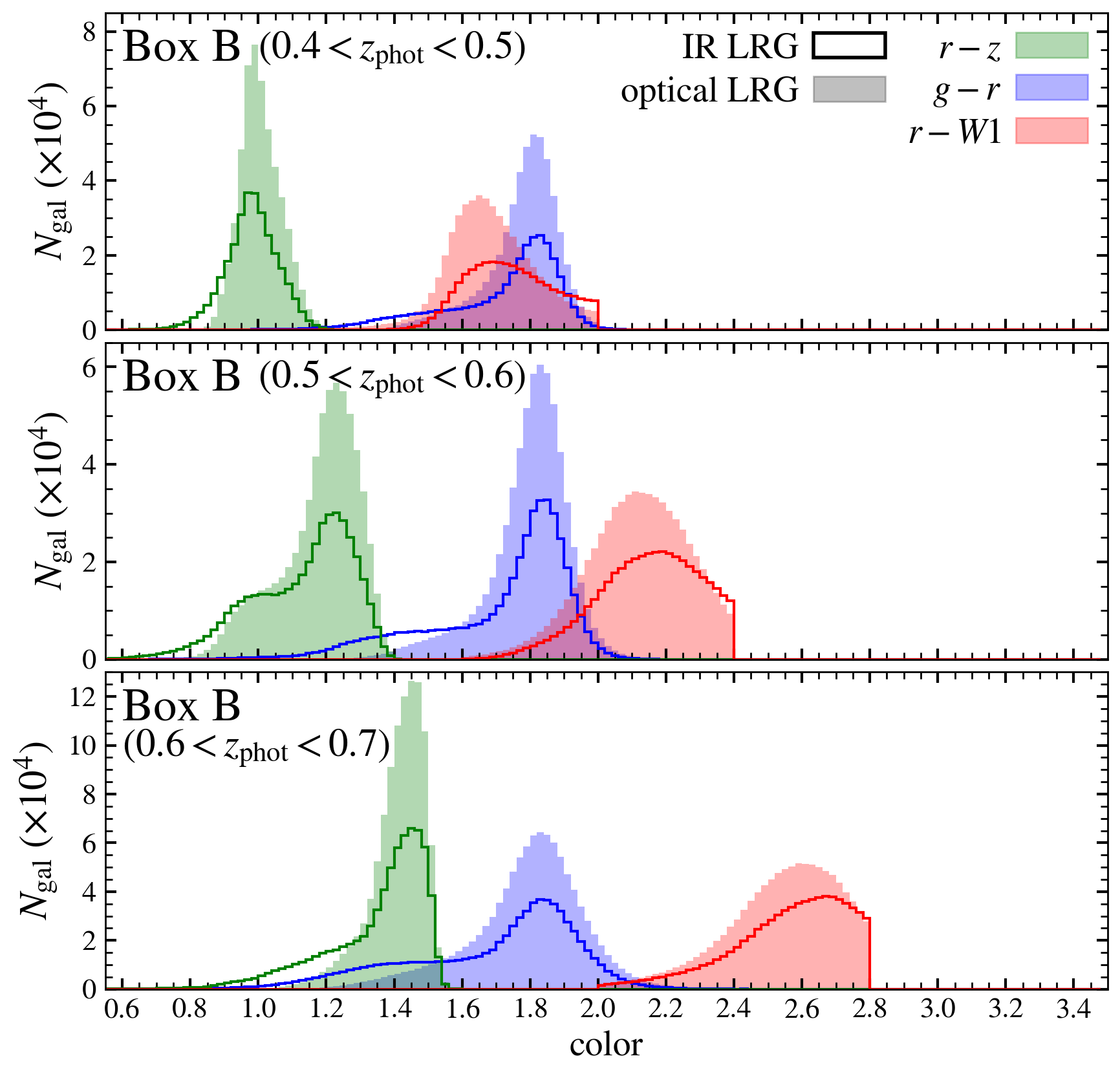}
  \caption{Distributions of ${r-z}$, ${g-r}$, and ${r-W1}$ colors of optical (shaded histograms) and IR (open histograms) LRGs from above (Box A; top three panels) and along (Box B; bottom three panels) the red sequence in ${r-W1}$ versus $M_{W1}$ color--magnitude space. The regions of color--magnitude space corresponding to Box A and Box B are shown in Figure~\ref{fig:cmd}. The mean, median, and skewness of each distribution, as well as the number of LRGs in each population, are given in Table~\ref{tab:data_sed}. Optical and IR LRGs from along the red sequence have similar distributions of each color, while optical LRGs above the red sequence have redder ${r-z}$ and ${g-r}$ and bluer ${r-W1}$ colors compared to IR LRGs in the same $W1$-band luminosity range.
  }
\label{fig:data_sed}
\end{figure}

\begin{deluxetable*}{ c }
\tablecaption{Statistics of distributions of ${r-z}$, ${g-r}$, and ${r-W1}$ colors of optical and IR LRGs selected from above (Box A) and along (Box B) the red sequence in IR color--magnitude space (${r-W1}$ versus $M_{W1}$). The boxed regions A and B are shown for the $0.4 < \zphot < 0.5$ redshift bin in Figure~\ref{fig:cmd}, and the color distributions are shown in Figure~\ref{fig:data_sed}. The mean, median, and skewness of each distribution is shown for IR (optical) LRGs in bold (plain) text.
\label{tab:data_sed}
}
\startdata
\includegraphics[width=0.98\linewidth]{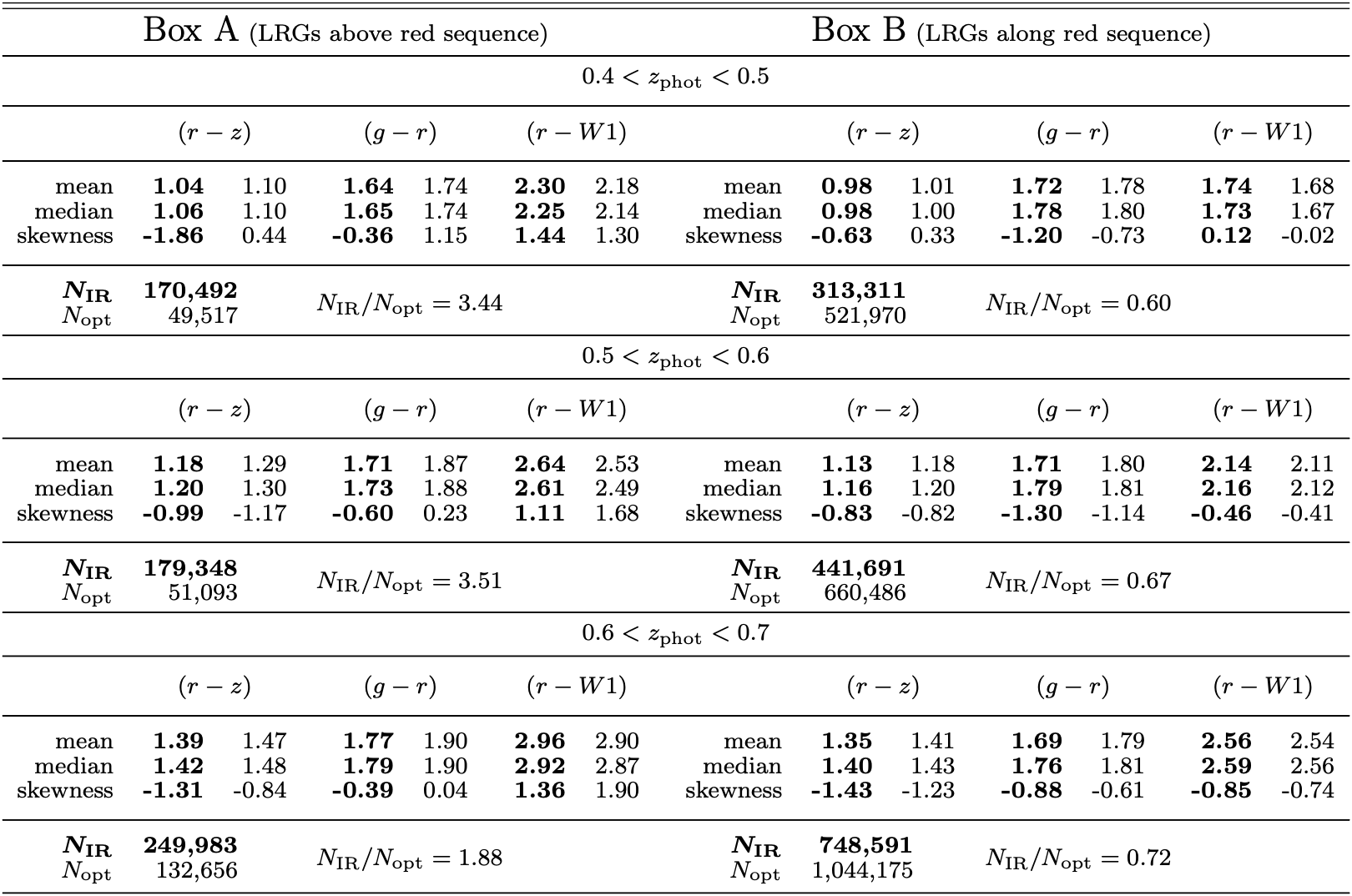}
\enddata
\end{deluxetable*}

Our $z$-band and $W1$-band models yield magnitude-limited mock galaxy catalogs from which we select mock LRGs. From these results we predict the clustering of DESI LRG target samples.

Figure~\ref{fig:wp_lrg} shows the predicted clustering of IR and optical mock LRGs compared to the relevant DESI LRG target sample. In addition to the data (black points with error bars), each panel of Figure~\ref{fig:wp_lrg} shows the clustering of mock LRGs according to the $z$-band and $W1$-band models, with and without the addition of scatter to the model color--\zstarve relation (\S\ref{subsec:color_assign}).

As shown in the top row of Figure~\ref{fig:wp_lrg}, the $z$-band and $W1$-band models predict very similar clustering signals for IR LRGs. The predicted amplitude of both the one-halo and two-halo terms exceeds that of the data in the two lowest redshift bins, and is in better agreement in the highest redshift bin ($\zsim=0.628$). Across all redshift bins the addition of scatter to the color--\zstarve relation brings the predicted clustering amplitude of IR LRGs closer to the data, reducing the clustering amplitude of the one-halo term by $\sim15$--$20\%$, and the two-halo term by $\sim10\%$, depending on redshift bin. However, this reduction in the discrepancy between the model and data is a ceiling achieved only by adding such a high degree of scatter that color is uncorrelated with our proxy for halo assembly history\footnote{We also tested using proxies for IR and optical LRGs by selecting the $N$ brightest mock galaxies from the relevant magnitude-limited mock galaxy catalog to recreate the number densities of the IR and optical DESI LRG target samples, instead of applying the DESI LRG target selections to the magnitude-limited mocks as described in \S\ref{subsec:mock_lrg_select}. The result was a slightly larger predicted clustering amplitude compared to the ``random colors" models that do use the DESI LRG target selections, shown in Figure~\ref{fig:wp_lrg}.} (\zstarve).

The bottom row of Figure~\ref{fig:wp_lrg} shows that the $z$-band and $W1$-band models predict different clustering amplitudes for optical LRGs. The $z$-band model overpredicts the clustering amplitude of optical LRGs to a similar degree as the overprediction for IR LRGs. Assigning colors at random such that galaxy color is uncorrelated with halo assembly history again reduces this discrepancy of the one-halo (two-halo) term by $\sim20$--$25\%$ ($\sim10$--$15\%$).

On the other hand, the $W1$-band model prediction for the clustering of optical LRGs is in very good agreement with the data, especially the two-halo term (the one-halo term is actually slightly underpredicted). Additionally, adding \emph{any} amount of scatter to the color--\zstarve relation has no effect on the predicted clustering amplitude in this case.

The distributions of optical and IR LRGs in color--magnitude space are helpful for interpreting Figure~\ref{fig:wp_lrg}. As Figure~\ref{fig:cmd} shows, in optical space (${M_r-M_z}$ versus $M_z$), both optical and IR LRGs are largely confined to the red sequence, with the distribution extending below it toward bluer color, especially at higher redshift.

In contrast, in IR space (${M_r-M_{W1}}$ versus $M_{W1}$) the galaxy distribution does extend significantly above the red sequence. For example, in the ${0.4 < \zphot < 0.5}$ redshift bin, the red sequence is at ${M_r-M_{W1}\sim1.6}$, but the distribution extends to ${M_r-M_{W1} \gtrsim 3.0}$. IR LRGs occupy the entire region of excess ${M_r-M_{W1}}$ color across the full range of $M_{W1}$, while optical LRGs occupy only the part of this region that corresponds to more luminous $M_{W1}$, i.e., the optical LRG selection excludes a population of galaxies with very red ${M_r-M_{W1}}$ colors and moderately luminous $W1$-band luminosities that are included by the IR selection.

The top panels of Figure~\ref{fig:data_sed} explore the differences in the color distributions of optical and IR LRGs located above the red sequence (Box A in Figure~\ref{fig:cmd}) in ${r-W1}$ versus $M_{W1}$ color--magnitude space. The bottom three panels (Box B in Figure~\ref{fig:cmd}) show the color distributions of optical and IR LRGs from the same $M_{W1}$ range \emph{along} the red sequence. 

Along the infrared red sequence, the SEDs of optical and IR LRGs are quite similar. However, the two selections diverge above the red sequence. Table~\ref{tab:data_sed} lists the mean, median, and skewness of each color distribution, as well as the number of LRGs in each population.

Above the red sequence IR LRGs have bluer ${r-z}$ and ${g-r}$ colors than optical LRGs in the same region of color--magnitude space, while IR LRGs have redder ${r-W1}$ colors than their optical counterparts from the same color--magnitude region.
The same general trend is seen along the red sequence, but the color differences between the optical and IR LRG selections are smaller, i.e., along the red sequence IR LRGs also have bluer ${r-z}$ and ${g-r}$ colors than optical LRGs, but by only half as much as above the red sequence. Similarly, IR LRGs along the red sequence have redder ${r-W1}$ colors than their optical counterparts, but the color difference is two to three times smaller than for IR and optical LRGs above the red sequence.

Activity from active galactic nuclei (AGN) may artificially inflate the observed IR ($W1$-band) luminosities and artificially redden the $r-W1$ colors of some galaxies that pass the IR LRG selection \citep[e.g.,][]{webster_etal95, georgakakis_etal09, banerji_etal12, glikman_etal12, kim_im18, klindt_etal19, rivera_etal21}, relative to galaxies with comparable IR luminosities and ${r-W1}$ colors but no AGN activity.
The bluer ${r-z}$ colors of IR LRG targets cause these objects to be excluded by the optical LRG selection (e.g., panel (e) of Figure~\ref{fig:not-lrg_outliers}).
Our IR ($W1$-band) model would assign mock IR LRGs with artificially red $r-W1$ colors and high IR luminosities due to AGN activity to halos with higher bias than if their $r-W1$ colors and $W1$-band magnitudes were not influenced by AGN activity.
This could explain why our $W1$-band model overpredicts the clustering amplitude of IR LRGs relative to optical LRGs, as shown in Figure~\ref{fig:wp_lrg}.

The influence of AGN activity may also explain the lack of correlation between IR and optical color when assigning colors to mock galaxies based solely on a proxy for halo age (here \zstarve). This modeling assumption attributes galaxy color entirely to halo mass accretion history, and does not account for how baryonic effects such as AGN activity may contribute to galaxy color. In other words, optical color is not necessarily a reliable proxy for IR color. This is also illustrated by Figure~\ref{fig:color_color}, which shows optical (${r-z}$) versus IR (${r-W1}$) color for both IR and optical LRGs in the $0.4 < \zphot < 0.5$ redshift bin. The optical and IR colors of optical LRGs are more closely correlated than for IR LRGs, but in both cases there is considerable scatter in $r-W1$ at fixed $r-z$.
\citet{xu_etal22} also find a lack of correlation between halo assembly properties and galaxy color using a semi-analytic model.

\subsection{The LRG--halo connection}\label{subsec:lrg-halo}

\begin{figure*}
\centering
    \includegraphics[width=\linewidth]{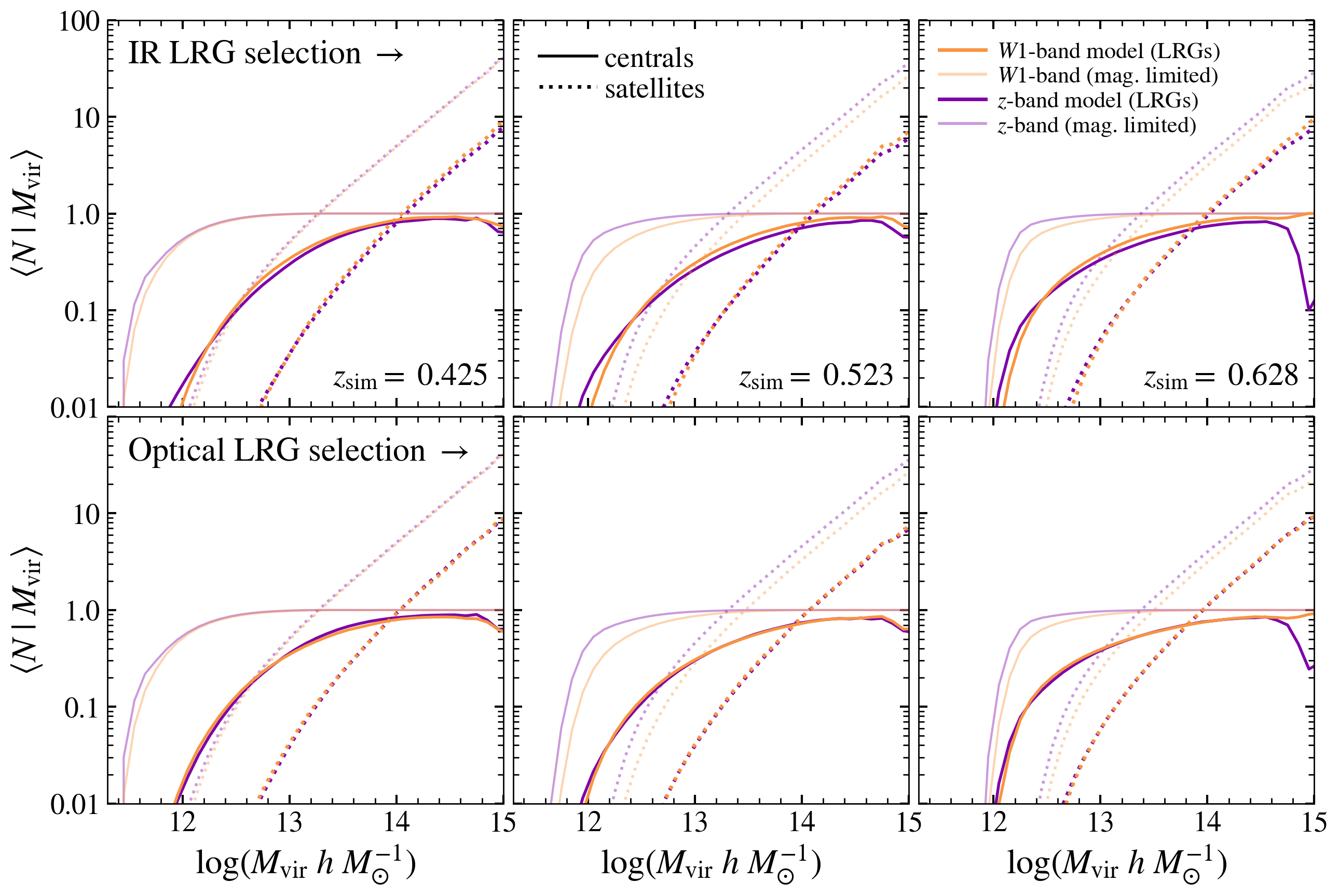}
    \caption{Predicted HODs of IR (top row) and optical (bottom row) mock LRGs from both the $z$-band (heavier purple lines) and $W1$-band (heavier orange lines) models. Also shown in fainter purple and orange lines are HODs of the $z$-band and $W1$-band magnitude-limited mock galaxy catalogs from which mock LRGs are selected. Solid (dotted) lines show results for central (satellite) galaxies and LRGs.
    Each column shows a different redshift bin.
    The best-fit values of a standard five-parameter HOD model, as well as the satellite fraction, for each mock LRG sample are given in Table~\ref{tab:hod_params}.
    }
\label{fig:hod_lrg}
\end{figure*}

\begin{deluxetable*}{ c c c c c }
\tablecaption{Predicted halo occupation completeness for mock central LRGs by selection (IR or optical). Results for the $W1$-band ($z$-band) model are in bold (plain) text.
\label{tab:hod_stats}
}
\tablehead{
\colhead{\zsim} & \colhead{LRG selection} &
\colhead{peak $\langle N_{\rm cen} | \mvir \rangle$\tablenotemark{a}} &
\colhead{min \mvir of peak $\langle N_{\rm cen} | \mvir \rangle$\tablenotemark{b}} &
\colhead{$\langle N_{\rm cen} | (\mvir > 10^{14.75}\ \msunh) \rangle$\tablenotemark{c}}
}
\startdata
\multirow{2}{*}{$0.425$} & IR & {\bf 0.92}~~0.91 & {\bf 14.45}~~14.85 & {\bf 0.74}~~0.70 \\
	& optical & {\bf 0.85}~~0.90 & {\bf 14.35}~~14.55 & {\bf 0.62}~~0.71 \\
\vspace{-1ex} \\
\hline
\vspace{-1ex} \\
\multirow{2}{*}{$0.523$} & IR & {\bf 0.91}~~0.87 & {\bf 14.45}~~14.65 & {\bf 0.77}~~0.61 \\
	& optical & {\bf 0.85}~~0.84 & {\bf 14.75}~~14.65 & {\bf 0.69}~~0.64 \\
\vspace{-1ex} \\
\hline
\vspace{-1ex} \\
\multirow{2}{*}{$0.628$} & IR & {\bf 0.96}~~0.82 & {\bf 14.85}~~14.55 & {\bf 0.96}~~0.31 \\
	& optical & {\bf 0.92}~~0.84 & {\bf 14.95}~~14.55 & {\bf 0.92}~~0.38 \\
\enddata
\tablenotetext{a}{Highest central occupation fraction reached across all halo \mvir.}
\tablenotetext{b}{Minimum halo \mvir at which the highest central occupation fraction is achieved.}
\tablenotetext{c}{Mean central occupation fraction at the largest halo masses ($\mvir \gtrsim 10^{14.75}\ \msunh$), which is universally less or equal to than the peak central occupation fraction.}
\end{deluxetable*}

We compute the mean central and satellite halo occupation statistics as a function of halo mass directly from the mock galaxy catalogs and halo abundances from the MDPL2 simulation. The results are shown in Figure~\ref{fig:hod_lrg}.
Each panel of Figure~\ref{fig:hod_lrg} shows ${\langle N | \mvir \rangle}$ for the full magnitude-limited mock galaxy catalog in gray, while optical and IR LRGs are shown in purple and orange, respectively.

The key conclusions of Figure~\ref{fig:hod_lrg} are summarized in Table~\ref{tab:hod_stats}, which gives the peak value of ${\langle N_{\rm cen} \rangle}$ for optical and IR LRGs predicted by both the $z$-band and $W1$-band models, as well as the mean value of ${\langle N_{\rm cen} \rangle}$ for the most massive halos, which is universally less than the corresponding peak value.

The $z$-band model does not predict significant differences between the populations selected by the IR and optical LRG selection functions. According to this model the central LRG halo occupation fraction peaks at $\sim90\%$ at $z\sim0.43$, and drops to $\sim70\%$ at the largest halo masses, while by $z\sim0.63$ $\langle N_{\rm cen}\rangle$ peaks at 82--84\% and falls to $\sim35\%$ for the most massive halos, regardless of LRG selection function.

In contrast, the $W1$-band model \emph{does} predict a different LRG--halo relationship for IR versus optical LRGs. In all redshift bins this model has ${\langle N_{\rm cen} \rangle}$ for IR LRGs peaking at around 92\% and remaining at $\gtrsim75\%$ for the most massive halos, while ${\langle N_{\rm cen} \rangle}$ for optical LRGs peaks at 85--92\%, and falls at the largest halo masses as low as 62\% ($z\sim0.43$) to 69\% ($z\sim0.52$), although it remains at 92\% for the highest redshift bin ($z\sim0.63$).

\begin{deluxetable}{ l r }
\tablecaption{Prior ranges for HOD fit parameters (Eqs.~\ref{eq:hod_cen} and \ref{eq:hod_sat}).
\label{tab:hod_params_priors}
}
\tablehead{
\colhead{Parameter} & \colhead{Prior interval}
}
\startdata
$\log(M_{\rm min})$ & $(11.0,\ 14.0)$ \\
$\sigma_{\log M}$ & $(0.001,\ 1.5)$ \\
$\alpha$ & $(0.0,\ 2.0)$ \\
$\log(M_0)$ & $(11.0,\ 14.0)$ \\
$\log(M_1)$ & $(11.5,\ 15.5)$ \\
\enddata
\end{deluxetable}

\begin{deluxetable*}{ c c c c c c c c }
\tablecaption{Best-fit values of standard five-parameter HOD model (Eqs.~\ref{eq:hod_cen} and \ref{eq:hod_sat}), and the satellite fraction, $f_{\rm sat}$, of IR and optical mock LRG samples. Results for the $W1$-band ($z$-band) model are in bold (plain) text.
\label{tab:hod_params}
}
\tablehead{
\colhead{\zsim} & \colhead{LRG selection} &
\colhead{$\log(M_{\rm min}/\msunh)$} &
\colhead{$\sigma_{\log M}$} & \colhead{$\alpha$} &
\colhead{$\log(M_0/\msunh)$} &
\colhead{$\log(M_1/\msunh)$} &
\colhead{$f_{\rm sat}$}
}
\startdata
\multirow{2}{*}{0.425} & IR & {\bf 13.28}~~13.35 & {\bf 0.93}~~0.95 & {\bf 0.98}~~1.00 & {\bf 12.93}~~12.87 & {\bf 14.02}~~14.07 & {\bf 14.8}~~14.5 \\
 & optical & {\bf 13.32}~~13.26 & {\bf 1.05}~~0.93 & {\bf 0.95}~~0.98 & {\bf 12.91}~~12.89 & {\bf 14.00}~~14.00 & {\bf 14.9}~~14.8 \\
\vspace{-1ex} \\
\hline
\vspace{-1ex} \\
\multirow{2}{*}{0.523} & IR & {\bf 13.37}~~13.52 & {\bf 0.99}~~1.14 & {\bf 0.94}~~0.91 & {\bf 12.94}~~12.98 & {\bf 14.05}~~14.09 & {\bf 14.4}~~14.5 \\
 & optical & {\bf 13.45}~~13.44 & {\bf 1.15}~~1.11 & {\bf 0.95}~~0.96 & {\bf 12.84}~~12.85 & {\bf 14.05}~~14.05 & {\bf 14.7}~~14.6 \\
\vspace{-1ex} \\
\hline
\vspace{-1ex} \\
\multirow{2}{*}{0.628} & IR & {\bf 13.26}~~13.40 & {\bf 1.02}~~1.18 & {\bf 0.95}~~0.95 & {\bf 12.88}~~12.82 & {\bf 13.95}~~13.99 & {\bf 14.1}~~14.4 \\
 & optical & {\bf 13.30}~~13.31 & {\bf 1.19}~~1.16 & {\bf 0.96}~~0.94 & {\bf 12.78}~~12.82 & {\bf 13.93}~~13.93 & {\bf 14.4}~~14.7 \\
\enddata 
\end{deluxetable*}

We fit a standard five-parameter HOD model to each optical and IR mock LRG sample Table~\ref{tab:hod_params}.
For ease of comparison with their results we use the same functional form as \citet{zhou_etal20b}, who derive HOD parameters from clustering measurements of DESI LRG targets using photometric redshifts in bins that approximately match the redshift bins we use. This parameterization is described in detail in \citet{zheng_etal05} and \citet{zheng_etal07}.
Briefly, the probability $\langle N_{\rm cen} \rangle$ that a halo of virial mass \mvir hosts a central galaxy is given by
\begin{equation}\label{eq:hod_cen}
  \langle N_{\rm cen} | \mvir \rangle = \frac{1}{2} \! \left( 1 + {\rm erf}{\left[\frac{\log(\mvir)-\log(M_{\rm min})}{\sigma_{\log M}} \right]} \right),
\end{equation}
\noindent where the parameter $\log(M_{\rm min})$ is the minimum halo mass for hosting a central galaxy, and the parameter $\sigma_{\log M}$ defines the steepness of the transition of $\langle N_{\rm cen} \rangle$ from $\sim0$ at low \mvir to $\sim1$ at high \mvir.

The mean number of satellite galaxies, $\langle N_{\rm sat} \rangle$, hosted by a halo of virial mass \mvir is approximated by a power law given by 
\begin{equation}\label{eq:hod_sat}
  \langle N_{\rm sat}|\mvir\rangle = {\left(\frac{\mvir-M_0}{M_1} \right)}^{\alpha},
\end{equation}
\noindent where $M_0$, $M_1$, and $\alpha$ are the remaining free parameters of the HOD fit. The HOD fits of \citet{zhou_etal20b} incorporate a sixth nuisance parameter to account for photometric redshift uncertainty, which in this work is addressed by the magnitude-dependent line-of-sight scatter parameter, \sigmalos (Eq.\ \ref{eq:sigmalos}).
Table~\ref{tab:hod_params_priors} gives the prior interval for each HOD parameter. The best-fit HOD parameters and satellite fraction of each mock LRG sample are given in Table~\ref{tab:hod_params}.

\subsection{Comparison with other ``DESI-like" LRG studies}\label{subsec:lit_compare}

\begin{figure*}
\centering
  \includegraphics[width=0.7\linewidth]{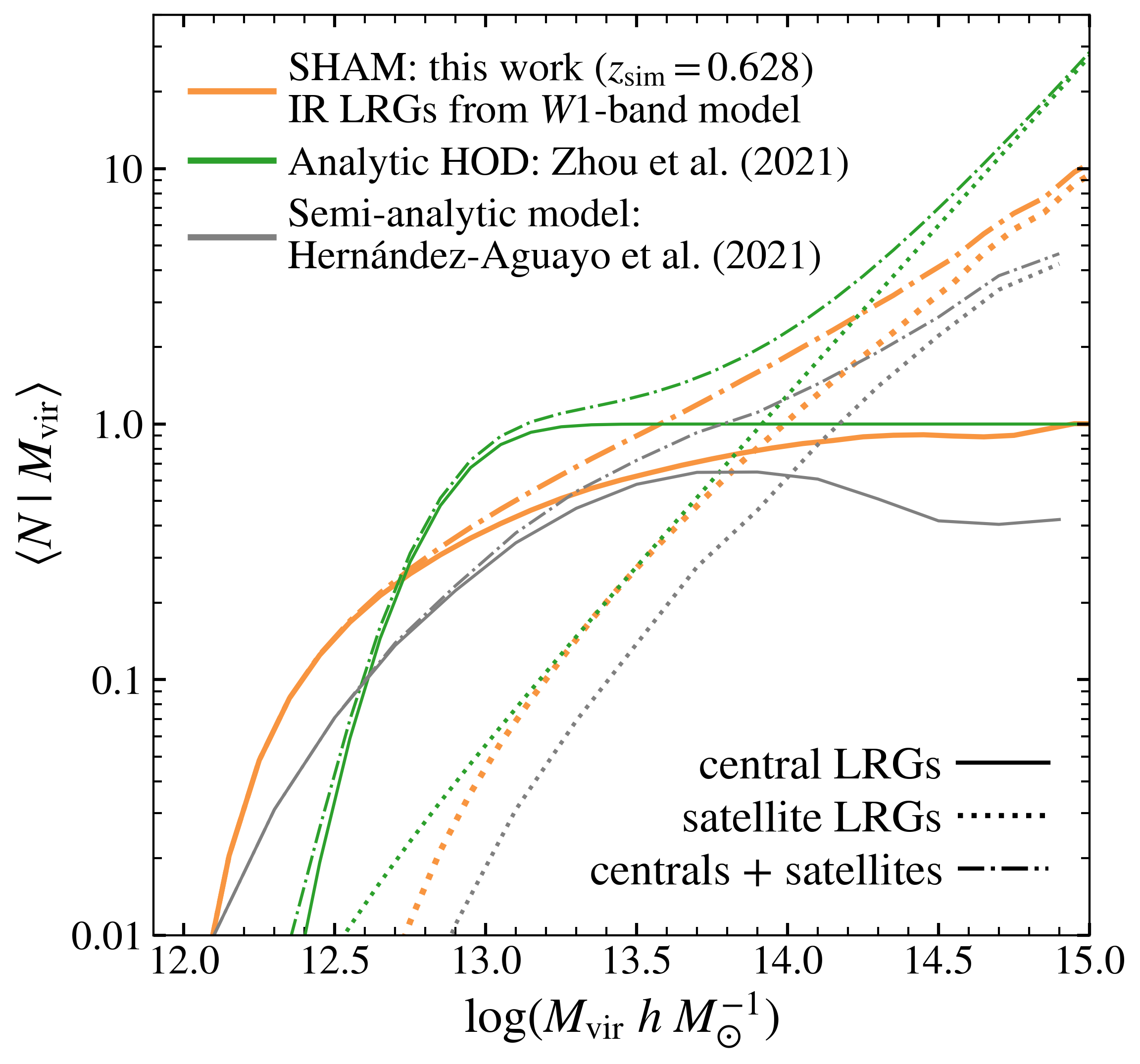}
  \caption{Comparison of LRG HOD parameters from this work at $z\sim0.63$ (orange lines) with the analytic HOD fits of \citet{zhou_etal20b} (green lines) and semi-analytic model of \citet{hernandez-aguayo_etal21} (gray lines) in comparable redshift bins.
  Solid, dotted, and dash-dotted lines show results respectively for central LRGs, satellite LRGs, and the combination of both.
  }
  \label{fig:hod_compare}
\end{figure*}

There are two main studies of the LRG--halo connection of ``DESI-like" LRGs suitable for comparison with our results. \citet{zhou_etal20b} (who provide the photometric redshifts used here) measure the clustering of LRGs at ${0.41 < \zphot < 0.93}$ selected from DECaLS DR7 photometry using the optical target selection (Eq.\ \ref{eq:lrg_opt}), and fit a standard five-parameter analytic HOD model in five redshift bins. 

\citet{zhou_etal20b} find little to no evolution of HOD parameters for LRGs across ${0.4 \lesssim z \lesssim 0.8}$. Our results are consistent with a lack of HOD parameter evolution across the subset of this redshift range that we model, although some of our predicted parameter values themselves differ from those of \citet{zhou_etal20b}, particularly $\sigma_{\log M}$. \citet{zhou_etal20b} find a very steep transition from a ${\langle N{\rm cen} \rangle \sim0}$ to 1 ($\sigma_{\log M}\sim0$--0.28; see their Figure 13), while our models predict a more gradual transition, with ${\sigma_{\log M}\sim1}$.

Figure~\ref{fig:hod_compare} compares our predicted LRG halo occupation statistics at $z\sim0.63$ with the best-fit HOD parameterization of \citet{zhou_etal20b} in the relevant redshift bin from their study ($0.61 < \zphot < 0.72$). Of the redshift bins \citet{zhou_etal20b} use that overlap with this work, this is the bin in which they find the greatest deviation of $\langle N{\rm cen} \rangle$ from a step function, although we also select this bin for Figure~\ref{fig:hod_compare} to enable comparison with an additional study, discussed later in this section.

Our HOD predictions for satellite LRGs agree with those of \citet{zhou_etal20b} at halo masses below ${\log(\mvir/ \msunh) \sim 14}$, and our overall satellite fractions of $f_{\rm sat}\sim0.14$--0.15 across both model bands and LRG selections is consistent with \citet{zhou_etal20b}, who find $f_{\rm sat}\sim0.13$--0.16, depending on redshift.

\citet{zhou_etal20b} obtain best-fit parameters for analytic forms of central and satellite LRG HODs (Eqs.\ \ref{eq:hod_cen} and \ref{eq:hod_sat}), which will necessarily correspond to 100\% of massive halos containing a central LRG due to the functional form of Eq.\ \ref{eq:hod_cen} (this is also the case for analytic HOD fits to our mock LRG samples, shown in Table~\ref{tab:hod_params}). Enforcing 100\% halo occupation at the high-mass end by LRGs also suppresses accounting for potential contamination by lower-mass halos. 

As Figure~\ref{fig:hod_lrg} shows, our models do \emph{not} predict that central LRG halo occupation reaches unity, instead peaking at 82--96\% and falling in most cases to $\sim60$--77\% for the most massive halos (${\log(\mvir/ \msunh) \gtrsim 14,75}$), although this depends on both model band ($z$ or $W1$) and LRG selection (IR or optical; see Table~\ref{tab:hod_stats}).

The other study to which we can compare our results is \citet{hernandez-aguayo_etal21}, who use the \textsc{galform} semi-analytic galaxy formation model \citep{cole_etal00} and nine snapshots from the Planck-Millennium $N$-body simulation \citep{baugh_etal19} to study the galaxy--halo connection of ``DESI-like" LRGs at redshifts between 0.6 and 1.0. Their LRG HOD results for $z=0.64$ are overlaid with our $\zsim\sim0.63$ results in Figure~\ref{fig:hod_compare}.

\citet{hernandez-aguayo_etal21} predict central and satellite LRG abundances below our model predictions, although they note that \textsc{galform} underpredicts the abundance of LRGs across the entire redshift range they study, with the greatest discrepancy at $z\sim0.6$--0.7.
The shape of their central LRG HOD is notably similar to ours, consistent with a gradual transition to a peak central LRG halo occupation less than unity that decreases at the largest halo masses. Specifically, \citet{hernandez-aguayo_etal21} find that ${\langle N_{\rm cen}|\mvir\rangle}$ for ``DESI-like" LRGs reaches a maximum of $\sim70\%$ at ${\log(\mvir/\msunh) \sim 13.75}$, and is as low as $\sim40\%$ at ${\log(\mvir/\msunh) \sim 14.5}$.

\section{Summary and conclusion}\label{sec:summary}
We have used a SHAM and age distribution matching framework to construct magnitude-limited mock galaxy catalogs at $z\sim0.43$, 0.52, and 0.63 with ${r-z}$ and ${r-W1}$ colors. From these catalogs we select mock LRG samples according to both the optical and IR DESI LRG target selection functions.
This work is the first application SHAM modeling in the infrared, complimenting the few existing studies of the galaxy--halo connection of ``DESI-like" LRGs.

Our models reproduce the number densities, luminosity functions, color distributions, and magnitude-dependent projected clustering (in the luminosity range of DESI LRG targets) of the parent galaxy samples from the Legacy Surveys DR9 photometry that serve as the basis for DESI LRG target selection. With the mock LRG samples selected from our magnitude-limited mock galaxy catalogs, we predict the halo occupation statistics of both optical and IR DESI LRGs at a fixed cosmology. We assess the differences between these two LRG populations, as well as the effect of using the $z$-band versus the 3.4 micron $W1$-band for SHAM and age distribution matching.

The main results of this work are:

\begin{enumerate}
  \item Both the optical and IR DESI LRG target selections exclude some of the most luminous galaxies that would appear to be LRGs based on their position on the red sequence in optical color--magnitude space (Figure~\ref{fig:lrg_frac}). This is a result of the specific DESI LRG target selection cut intended to exclude blue, low-redshift (${z \lesssim 0.4}$) galaxies (Figure~\ref{fig:not-lrg_outliers}, \S\ref{subsec:mock_lrg_select}).

  \item Optical and IR LRGs occupy similar regions of optical color--magnitude space (${r-z}$ versus $M_z$), where the red sequence corresponds to the very reddest ${r-z}$ colors (Figure~\ref{fig:cmd}). In IR color--magnitude space (${r-W1}$ versus $M_{W1}$) there is a sizable galaxy population at significantly redder colors than the red sequence. IR LRGs occupy most of this region of excess ${r-W1}$ color, but optical LRGs are largely excluded. This could be a result of AGN activity artificially inflating the $W1$-band luminosities for some of these objects (\S\ref{subsec:clust_lrg}).
  
  \item There are clear distinctions between the LRG samples obtained from the optical versus IR selections that are apparent from the data alone, namely among their optical ($r-z$ versus $M_z$) and IR ($r-W1$ versus $M_{W1}$) color--magnitude distributions, clustering, and color--color distributions. Our IR-based ($W1$-band) model predicts greater differences between the halo occupation statistics of optical and IR LRGs than the $z$-band model (Figure~\ref{fig:hod_lrg}, Table~\ref{tab:hod_stats}), and is therefore the preferred regime for this comparative study of the two selections.

  \item Age distribution matching, which assumes a monotonic correlation between halo age and galaxy color at fixed luminosity, tends to overpredict the clustering amplitude of DESI LRGs (Figure~\ref{fig:wp_lrg}). Introducing scatter into the assumed age--color relation improves agreement between predictions and the data somewhat, although the greatest improvement comes from increasing this scatter to a level that is equivalent to assigning colors at random. Our models therefore suggest that either galaxy color is uncorrelated with halo age in the LRG regime, or there is some additional model parameter that has been neglected and is possibly related to our use of photometric redshifts.

 \item Both DESI LRG target selections yield populations with a non-trivial LRG--halo connection that does not reach unity ($\langle N_{\rm cen} \rangle \sim 1$) for the most massive halos (Figure~\ref{fig:hod_lrg}, \S\ref{subsec:lrg-halo}). However, the IR selection achieves greater completeness ($\langle N_{\rm cen} \rangle \geq 91\%$) than the optical selection ($\langle N_{\rm cen} \rangle \sim 85$--$92\%$) across all redshift bins studied. Our results of $\langle N_{\rm cen} \rangle < 1$ is qualitatively consistent with the SAM predictions of \citet{hernandez-aguayo_etal21} (Figure~\ref{fig:hod_compare}, \S\ref{subsec:lit_compare}), although they find a lower maximum completeness of $\sim70\%$ at $z\sim0.64$.
\end{enumerate}

A natural extension of this work would be to utilize spectroscopic redshifts from the DESI BGS sample \citep{hahn_etal22} to conduct a similar study at lower redshift. The BGS\footnote{Specifically the BGS Bright sample; BGS also contains BGS Faint and AGN samples with different selection criteria.} is nearing completion and is obtaining spectroscopic redshifts for a magnitude-limited ($r<19.5$) sample of over 800 galaxies per \degsq at $z < 0.4$.
While the LRG target selection algorithms used in this work are designed to identify LRGs at $0.4 \lesssim z \lesssim 1.0$, these cuts could easily be modified to select LRGs in the same redshift range as the BGS. Mock LRGs could then be selected from SHAM-based magnitude-limited mock galaxy catalogs tuned to recreate the magnitude and color distributions of the BGS sample.

\section*{Acknowledgements}
The authors thank Zheng Zheng for comments on an earlier draft of this work.

The Legacy Surveys consist of three individual and complementary projects:\ the Dark Energy Camera Legacy Survey (DECaLS; Proposal ID \#2014B-0404; PIs:\ David Schlegel and Arjun Dey), the Beijing-Arizona Sky Survey (BASS; NOAO Prop.\ ID \#2015A-0801; PIs:\ Zhou Xu and Xiaohui Fan), and the Mayall $z$-band Legacy Survey (MzLS; Prop.\ ID \#2016A-0453; PI:\ Arjun Dey). DECaLS, BASS and MzLS together include data obtained, respectively, at the Blanco telescope, Cerro Tololo Inter-American Observatory, NSF's NOIRLab; the Bok telescope, Steward Observatory, University of Arizona; and the Mayall telescope, Kitt Peak National Observatory, NOIRLab. The Legacy Surveys project is honored to be permitted to conduct astronomical research on Iolkam Du'ag (Kitt Peak), a mountain with particular significance to the Tohono O'odham Nation.

NOIRLab is operated by the Association of Universities for Research in Astronomy (AURA) under a cooperative agreement with the National Science Foundation.

This project used data obtained with the Dark Energy Camera (DECam), which was constructed by the Dark Energy Survey (DES) collaboration. Funding for the DES Projects has been provided by the U.S. Department of Energy, the U.S. National Science Foundation, the Ministry of Science and Education of Spain, the Science and Technology Facilities Council of the United Kingdom, the Higher Education Funding Council for England, the National Center for Supercomputing Applications at the University of Illinois at Urbana-Champaign, the Kavli Institute of Cosmological Physics at the University of Chicago, Center for Cosmology and Astro-Particle Physics at the Ohio State University, the Mitchell Institute for Fundamental Physics and Astronomy at Texas A\&M University, Financiadora de Estudos e Projetos, Fundacao Carlos Chagas Filho de Amparo, Financiadora de Estudos e Projetos, Fundacao Carlos Chagas Filho de Amparo a Pesquisa do Estado do Rio de Janeiro, Conselho Nacional de Desenvolvimento Cientifico e Tecnologico and the Ministerio da Ciencia, Tecnologia e Inovacao, the Deutsche Forschungsgemeinschaft and the Collaborating Institutions in the Dark Energy Survey. The Collaborating Institutions are Argonne National Laboratory, the University of California at Santa Cruz, the University of Cambridge, Centro de Investigaciones Energeticas, Medioambientales y Tecnologicas-Madrid, the University of Chicago, University College London, the DES-Brazil Consortium, the University of Edinburgh, the Eidgenossische Technische Hochschule (ETH) Zurich, Fermi National Accelerator Laboratory, the University of Illinois at Urbana-Champaign, the Institut de Ciencies de l'Espai (IEEC/CSIC), the Institut de Fisica d'Altes Energies, Lawrence Berkeley National Laboratory, the Ludwig Maximilians Universitat Munchen and the associated Excellence Cluster Universe, the University of Michigan, NSF's NOIRLab, the University of Nottingham, the Ohio State University, the University of Pennsylvania, the University of Portsmouth, SLAC National Accelerator Laboratory, Stanford University, the University of Sussex, and Texas A\&M University.

BASS is a key project of the Telescope Access Program (TAP), which has been funded by the National Astronomical Observatories of China, the Chinese Academy of Sciences (the Strategic Priority Research Program ``The Emergence of Cosmological Structures'' Grant \#XDB09000000), and the Special Fund for Astronomy from the Ministry of Finance. The BASS is also supported by the External Cooperation Program of Chinese Academy of Sciences (Grant \#114A11KYSB20160057), and Chinese National Natural Science Foundation (Grant \#11433005).

The Legacy Survey team makes use of data products from the Near-Earth Object Wide-field Infrared Survey Explorer (NEOWISE), which is a project of the Jet Propulsion Laboratory/California Institute of Technology. NEOWISE is funded by the National Aeronautics and Space Administration.

The Legacy Surveys imaging of the DESI footprint is supported by the Director, Office of Science, Office of High Energy Physics of the U.S. Department of Energy under Contract No.\ DE-AC02-05CH1123, by the National Energy Research Scientific Computing Center, a DOE Office of Science User Facility under the same contract; and by the U.S. National Science Foundation, Division of Astronomical Sciences under Contract No.\ AST-0950945 to NOAO.

The Photometric Redshifts for the Legacy Surveys (PRLS) catalog used in this paper was produced thanks to funding from the U.S. Department of Energy Office of Science, Office of High Energy Physics via grant DE-SC0007914.

The authors gratefully acknowledge the Gauss Centre for Supercomputing e.V.\ (www.gauss-centre.eu) and the Partnership for Advanced Supercomputing in Europe (PRACE, www.prace-ri.eu) for funding the MultiDark simulation project by providing computing time on the GCS Supercomputer SuperMUC at Leibniz Supercomputing Centre (LRZ, www.lrz.de).

Some of the results in this paper have been derived using the \texttt{healpy} and \texttt{HEALPix} package.

\bibliography{refs}

\end{document}